\documentclass[aip, amsmath,amssymb, reprint,]{revtex4-1}

\usepackage{amsmath}
\usepackage{graphicx}
\usepackage{multirow}
\usepackage{array}
\usepackage{color} 
\usepackage{booktabs}
\usepackage{amsfonts}
\usepackage{hyperref}
\usepackage{textcomp}
\usepackage{ulem}
\usepackage[T1]{fontenc}
\usepackage{mathptmx}

\begin{document}

\title {How cold is the junction of a millikelvin scanning tunnelling microscope?}

\author {Taner Esat}
\affiliation {Peter Gr\"{u}nberg Institut (PGI-3), Forschungszentrum J\"{u}lich, 52425 J\"{u}lich, Germany}
\affiliation{J\"ulich Aachen Research Alliance (JARA), Fundamentals of Future Information Technology, 52425 J\"ulich, Germany}
\author {Xiaosheng Yang}
\affiliation {Peter Gr\"{u}nberg Institut (PGI-3), Forschungszentrum J\"{u}lich, 52425 J\"{u}lich, Germany}
\affiliation{J\"ulich Aachen Research Alliance (JARA), Fundamentals of Future Information Technology, 52425 J\"ulich, Germany}
\author {Farhad Mustafayev}
\affiliation {Peter Gr\"{u}nberg Institut (PGI-3), Forschungszentrum J\"{u}lich, 52425 J\"{u}lich, Germany}
\affiliation{J\"ulich Aachen Research Alliance (JARA), Fundamentals of Future Information Technology, 52425 J\"ulich, Germany}
\author {Helmut Soltner}
\affiliation{Zentralinstitut für Engineering, Elektronik und Analytik (ZEA-1), Forschungszentrum J\"ulich, 52425 J\"ulich, Germany}\author{F. Stefan Tautz}
\affiliation {Peter Gr\"{u}nberg Institut (PGI-3), Forschungszentrum J\"{u}lich, 52425 J\"{u}lich, Germany}
\affiliation{J\"ulich Aachen Research Alliance (JARA), Fundamentals of Future Information Technology, 52425 J\"ulich, Germany}
\affiliation{Experimentalphysik IV A, RWTH Aachen University, 52074 Aachen, Germany}
\author {Ruslan Temirov}
\email[corresponding author: ]{r.temirov@fz-juelich.de}
\affiliation {Peter Gr\"{u}nberg Institut (PGI-3), Forschungszentrum J\"{u}lich, 52425 J\"{u}lich, Germany}
\affiliation{University of Cologne, Faculty of Mathematics and Natural Sciences, Institute of Physics II, 50937 Cologne, Germany}

\keywords{scanning tunnelling microscopy, scanning tunnelling spectroscopy, low temperatures, superconductivity, Josephson tunnelling}

\begin{abstract}

We employ a scanning tunnelling microscope (STM) cooled to millikelvin temperatures by an adiabatic demagnetization refrigerator (ADR) to perform scanning tunnelling spectroscopy (STS) on an atomically clean surface of Al(100) in a superconducting state using normal-metal and superconducting STM tips. Varying the ADR temperatures between 30 mK and 1.2 K, we show that the temperature of the STM junction $T$ is decoupled from the temperature of the surrounding environment $T_{\mathrm{env}}$. Simulating the STS data with the $P(E)$ theory, we determine that $T_{\mathrm{env}} \approx 1.5$ K, while the fitting of the superconducting gap spectrum yields the lowest $T=77$ mK.

\end{abstract}

\maketitle
\section{Introduction}

The quantum effect of electron tunnelling lies in the core of scanning tunnelling microscopy (STM). STM junctions operated in a highly controlled manner under ultra-high vacuum (UHV) and low-temperature conditions enable high-resolution imaging, and precise manipulation of surface nanostructures \cite{Wagner2015}. However, further explorations of quantum-coherent phenomena in such nanostructures \cite{Chen2022} may demand even better control over the STM junction and its environment because of the utmost sensitivity of quantum tunnelling to the microscopic properties of the tunnelling junction and its environment. Since the temperature is one of the most critical environmental parameters, STM setups operating at ultra-low temperatures, i.e. well below 1 K, are being developed \cite{Song2010,Misra2013, Assig2013,Roychowdhury2014,Machida2018,Balashov2018,VonAllworden2018,Wong2020,Schwenk2020,Esat2021,Fernandez-Lomana2021}. 

The motivation for conquering the millikelvin (mK) temperature range with STM is two-fold: First, the ultra-low temperatures stabilize emergent quantum ground states with exceptional properties \cite{Machida2019,Nuckolls2020,Kamber2020}. Second, upon cooling the junction, the width of the Fermi-Dirac distribution in its electrodes shrinks, increasing the energy resolution of scanning tunnelling spectroscopy (STS) and, in that way, also providing better access to low-energy excitations with longer lifetimes \cite{Song2010_1,Yazdani2016,Feldman2017}. 

The problem of determining and later controlling the STM junction temperature is, however, anything but trivial. An obvious reason is that a direct temperature measurement typically performed with macroscopic sensors is technically impossible in such a microscopic junction. Another complication is that the junction permanently exchanges energy with its environment. These energy exchange processes affect the tunnelling rates appreciably \cite{Grabert1992}, necessitating the characterization of the environmental temperature, too.

One indirect way of deducing the junction temperature is to perform an STS experiment, i.e., to measure the dependence of the tunnelling current $I$ on the applied bias voltage $V$. Most generally,

\begin{equation} 
I=e[\Gamma^{\ +}(V)-\Gamma^{\ -}(V)],
\label{I}
\end{equation}

\noindent where $e$ is the elementary charge and $\Gamma^{\ \pm}$ is the tunnelling rate in the direction along (against) the applied bias voltage. %By definition:  $\Gamma^{\ +}(V)=\Gamma^{\ -}(-V)$. 
For an STM junction interacting with its electromagnetic environment \cite{Grabert1992}, $\Gamma^{\ +}$ can be calculated from

\begin{equation} 
\begin{split}
\Gamma^{\ +}(V)=\frac{4\pi}{\hbar}\iint_{-\infty}^{\infty} dE dE' n_{\mathrm{T}}(E)n_{\mathrm{S}}(E' + eV)\times \\  f_{\mathrm{T}}(E)[1-f_{\mathrm{S}}(E' + eV)]|M(E,E'+eV)|^2P(E-E'),
\end{split}
\label{G}
\end{equation}

\noindent where $n_{\mathrm{T,S}}$ are the densities of states (DOS) and $f_{\mathrm{T,S}}(E)=1/[1+\mathrm{exp}(E/k_{\mathrm{B}}T_{\mathrm{T,S}})]$ are the Fermi-Dirac distribution functions of the tip (T) and sample (S) electrodes. $k_{\mathrm{B}}$ is the Boltzmann constant and $M$ is Bardeen's matrix element accounting for the overlap between the single-electron wavefunctions in the tip and the sample \cite{Bardeen1961, Voigtlaender2015,Gottlieb2006}. Apart from the last factor under the integral, Eq. \ref{G} is identical to the classical result by Bardeen for the tunnelling current in a junction isolated from the environment and connected to an ideal voltage source \cite{Bardeen1961, Grabert1992}. 

The environment enters Eq. \ref{G} in the form of the so-called $P(E)$ function, which describes the probability for the tunnelling electron to exchange the energy $E$ with the environment \cite{Grabert1992}. $P(E-E')$ substitutes the delta function $\delta(E-E')$ in Bardeen's original expression for the non-interacting junction, thus accounting for the fact that also inelastic tunnelling processes occur. Note that it is the thermal dependence of $f_{\mathrm{T,S}}(E)$ that allows extracting the temperature of the tip $T_{\mathrm{T}}$ and the sample $T_{\mathrm{S}}$ from the STS data if $n_{\mathrm{T,S}}(E)$ and $P(E)$ are known. 

The problem with $n_{\mathrm{T, S}}(E)$ is that it generally results from a complicated and often irregular atomic structure of the tip or sample, respectively. For evaluating $T_{\mathrm{T,S}}$, it is therefore desirable to use materials with a well-defined DOS, on which atomic defects have little influence. It turns out that the best class of materials that satisfy this condition are the Bardeen-Cooper-Schrieffer (BCS) superconductors with their DOS given by the well-known expression \cite{Tinkham1996}

\begin{equation} 
n^{\mathrm{BCS}}(E)=n_0\mathrm{Re}\bigg[ \frac{1}{\sqrt{E^2-\Delta^2}}\bigg],
\label{BCS}
\end{equation}

\noindent in which $n_0$ defines the DOS in the normal state and $\Delta$ is the width of the superconducting gap. $n^{\mathrm{BCS}}$ features two spectroscopic singularities -- the so-called quasiparticle peaks -- situated symmetrically around $E_{\mathrm{F}}$ at the lowest quasiparticle excitation energy $\Delta$. Since according to Eq. \ref{BCS} the quasiparticle peaks are spectroscopically very sharp, the effect of their temperature-dependent broadening can help the accurate evaluation of $T_{\mathrm{T,S}}$ in a mK STM junction.

In a tunnelling junction comprising a normal metal tip and a superconducting surface, when the temperature is substantially lower than the critical temperature of the superconducting transition $T_{\mathrm{c}}$, Eq. \ref{G} for $\Gamma^{\ +}(V)$ simplifies to

\begin{equation} 
\begin{split}
\Gamma^{\ +}(V)=\frac{1}{e^2R}\iint_{-\infty}^{\infty} dE dE' n_{\mathrm{S}}^{\mathrm{BCS}}(E'+eV)\times \\f_{\mathrm{T}}(E)P(E-E'),
\end{split}
\label{G_NIS}
\end{equation}

\noindent losing its dependence on  $f_{\mathrm{S}}(E)$ and hence $T_{\mathrm{S}}$. Also note that under the assumption $n_{\mathrm{T}}(E)=const$, and $M(E,E')=const$, in the relevant range of energies, they factor out and together with $n_0$ are absorbed into $R$, the high-bias, or normal-state resistance of the junction. According to Eq. \ref{G_NIS}, tunnelling between a normal-metal tip and a superconducting surface yields $T_{\mathrm{T}}$ if the $P(E)$ function is known. Conversely, tunnelling between a superconducting tip and a normal-metal surface yields $T_{\mathrm{S}}$. Interestingly, in a junction with both electrodes being superconducting the tunnelling current, i.e, the Josephson current of Cooper pairs, is completely independent of $T_{\mathrm{T,S}}$ if $T_{\mathrm{T,S}} \ll T_{\mathrm{c}} $, and is given by

\begin{equation} 
I(V)=\frac{\pi e E_\mathrm{J}}{\hbar}\big[ P(2eV)-P(-2eV) \big],
\label{I_JO}
\end{equation}

\noindent which thus provides the most direct way of determining the $P(E)$ function experimentally \cite{Grabert1992}. In Eq. \ref{I_JO} $E_\mathrm{J}=\hbar I_\mathrm{c}/2e$ is the Josephson energy and $I_\mathrm{c}$ the critical Josephson current.

The first attempt at determining $P(E)$ of an STM junction is due to Ast and coworkers \cite{Jack2015, Ast2016}. Applying the $P(E)$ theory as sketched in the Methods section, they successfully simulated the experimental STS data collected with a dilution refrigerator operated mK STM \cite{Assig2013} on various tunnelling junctions \cite{Jack2015,Ast2016,Jack2016,Jack2017,Senkpiel2020_1,Senkpiel2020,Senkpiel2022}. They also demonstrated that the $P(E)$ function causes broadening of the STS data and thus imposes a fundamental limitation on the resolution of STS experiments. Furthermore they proposed that the effective capacitance of the STM junction $C$ is an important factor determining the degree of that broadening \cite{Ast2016}. 

Up to now, the $P(E)$ theory has been applied to STM, under the assumption that the junction is well thermalized with its environment, or in other words, the temperature of the junction is equal to that of its environment: $T_\mathrm{T, S}=T_\mathrm{env}$. It is, however, possible to envision a situation where this condition does not hold \cite{Martinis1993, Siewert1997}. 

Here we demonstrate that the junction of our mK STM is much colder than the surrounding environment $T_{\mathrm{T}} \ll T_{\mathrm{env}}$ by performing variable temperature STS of the Al(100) surface in its superconducting state, using both normal-metal and superconducting STM tips. First, we measured the temperature-dependent Josephson conductance and recovered the $P(E)$ function of the environment together with its temperature $T_{\mathrm{env}}$ from these data. Then we employed the obtained $P(E)$ function and $T_{\mathrm{env}}$ for evaluating the temperature $T_{\mathrm{T}}$ of the STM tip, which we consider to be an upper estimate of the junction temperature. Specifically, we obtained $T_\mathrm{T}$ by fitting the STS of the superconducting gap measured with a normal metal tip. In doing so, we additionally expose a potential problem of the commonly used fitting approach and suggest how to avoid it. Finally, we argue that the comparatively warm environment of our mK STM junction results in photon-assisted tunnelling that is responsible for the in-gap conductance seen in the STS of the superconducting gap.

\begin{figure}
\centering
\includegraphics[width=8cm]{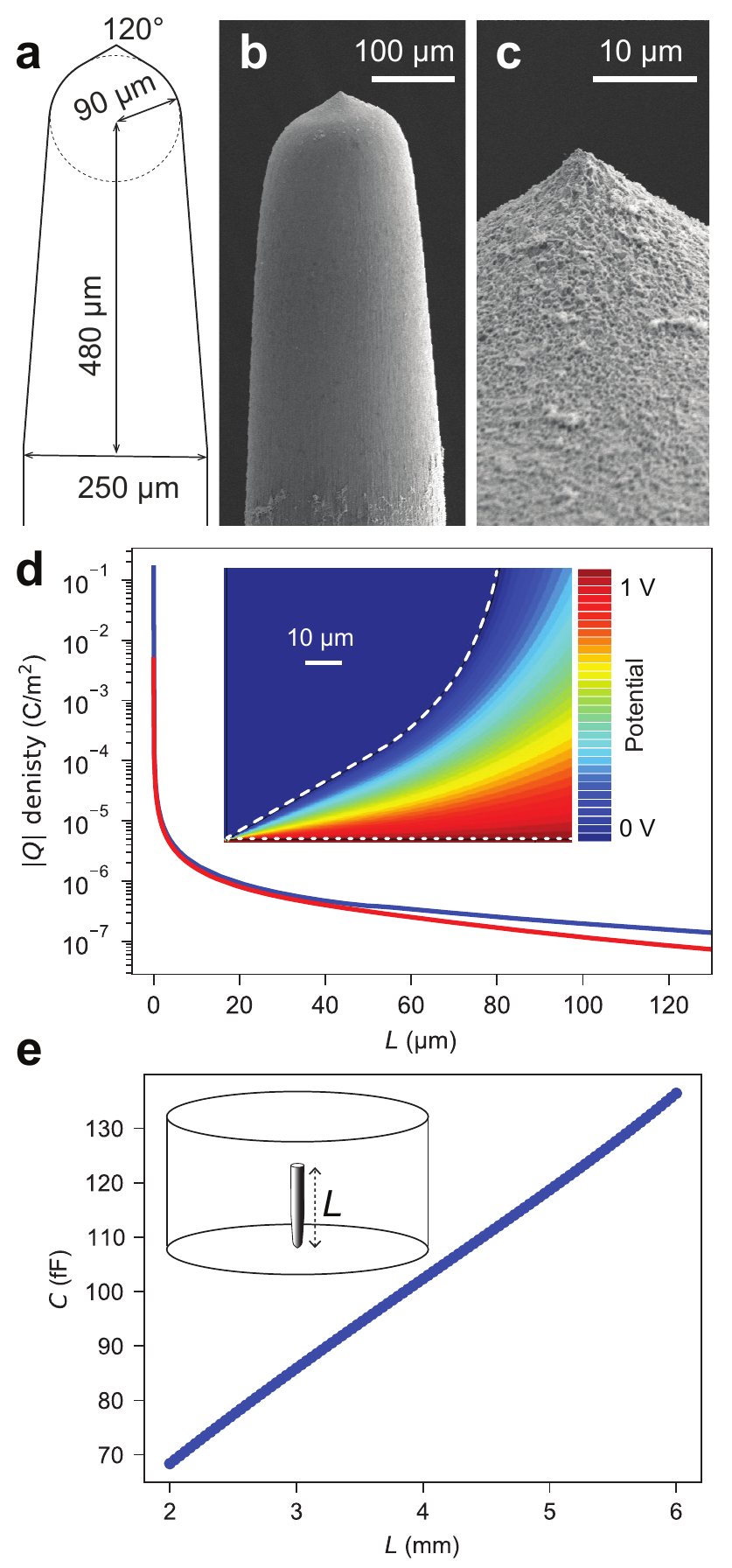}
\caption{The PtIr tip used in the experiments. (\textbf{a}) Model of the tip based on the SEM images (b-c). (\textbf{b-c}) SEM images of the tip. (\textbf{d}) Simulated surface charge density in the junction biased by the voltage of 1 V. (inset) Distribution of the potential in the junction. The dashed white line represents the outline of the tip. (\textbf{e}) Dependence of the junction capacitance $C$ on the length $L$ of the tip wire. (inset) Model of the STM head used for simulating $C$.}
\label{tip} 
\end{figure} 

\begin{figure}
\centering
\includegraphics[width=8cm]{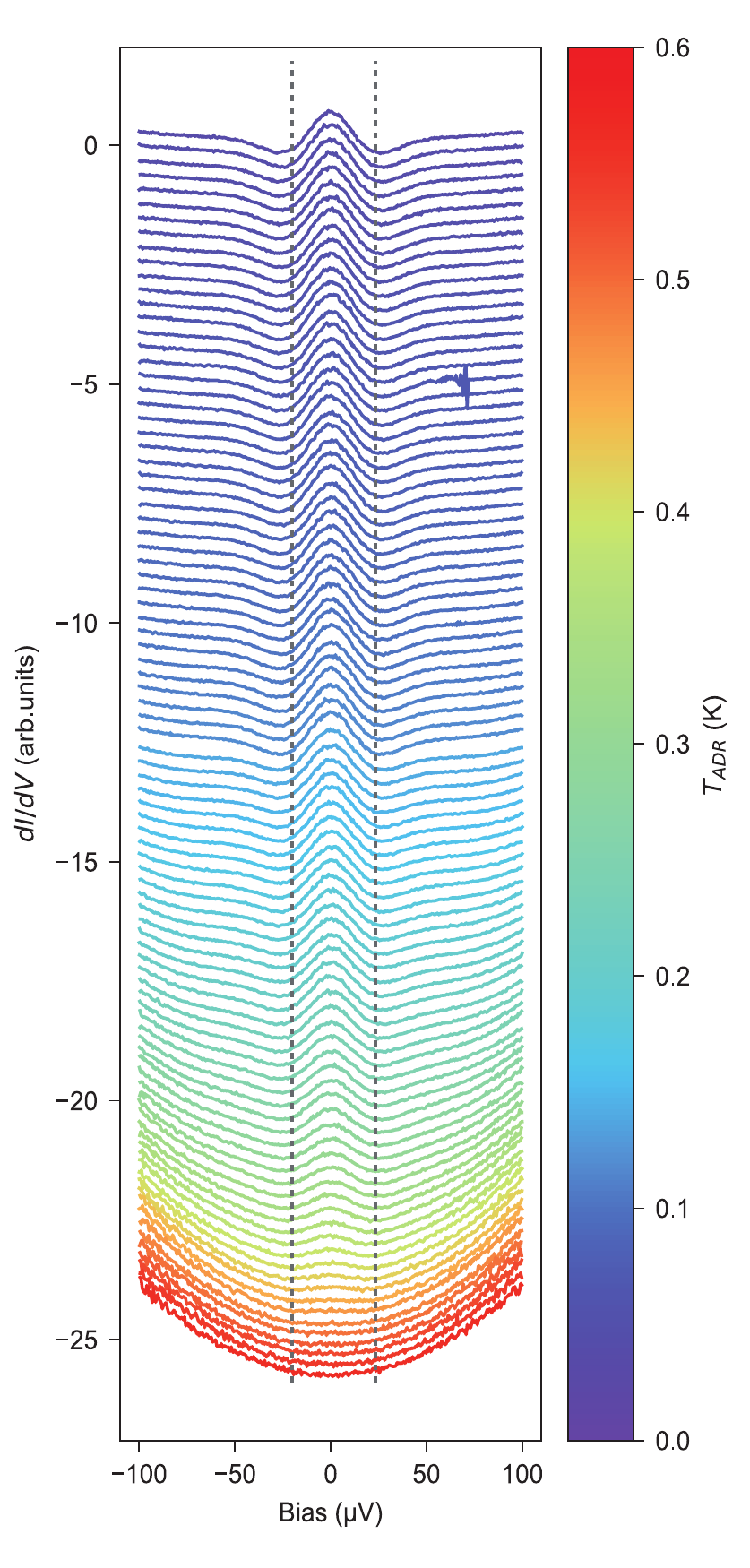}
\caption{Josephson conductance spectra measured with a superconducting tip (cf. text) on a superconducting Al(100) surface. Higher temperature data are shifted downwards for clarity. The dotted lines are eye guides exhibiting the temperature dependence of the Josephson conductance peak.}
\label{joseph} 
\end{figure}

\section{Results and discussion}

\subsection{Capacitance of the tip-sample junction}

Considering the broadening effect of the tip-surface junction capacitance $C$ on STS \cite{Ast2016}, we attempted to increase $C$ by using blunter PtIr tips. The model of the tip geometry exhibited in Fig. \ref{tip}a was extracted from the scanning electron microscopy images shown in Fig. \ref{tip}b-c: We approximated the tip apex shape by a sphere 90 $\mu$m in radius and crowned by a cone with 120$^\circ$ opening angle. 

Placing the model tip over a flat metal surface, we simulated the charge distribution in the tip-surface junction with a commercial software employing the boundary element method (BEM) \cite{BEM}. Fig. \ref{tip}d shows the simulated surface charge resulting from a potential difference of 1 V applied to the surface while the grounded tip is located at a distance of 1 nm. Although the induced charge density drops quickly with the distance from the tip apex, a more careful analysis shows that the density of charge that accumulates on the shaft of the tip, i.e., far from the apex, is not negligible.

Because the energy resolution of a mK STM may be as high as 10 $\mu$eV which corresponds to $\approx2.5$ GHz \cite{Schwenk2020}, the dimensions contributing to the junction capacitance may reach the scale of several cm. We, therefore, considered the complete STM head \cite{Esat2021} for evaluating $C$. A simplified model that captures the essential features of our STM head (Fig. \ref{tip}e) pictures the junction capacitor as a cylindrical metallic cavity with a radius of $7 \ \mathrm{mm}$ and a height of 7 mm inside which the tip, i.e., the second electrode, is located. The bottom of the cavity coincides with the surface plane of the sample; hence, the distance from the tip to the cavity bottom is 1 nm. Systematically changing the length of the tip wire $L$ between 2 and 6 mm, we obtained the plot shown in Fig. \ref{tip}e, which demonstrates that $C$ depends on $L$ and hence the correct evaluation of $C$ needs to account for such additional elements as the tip holder and possibly also the tunnelling current wire connected to the tip. Instead of performing more elaborate simulations, we pick $C=100$ fF as a ballpark value for our further analysis. 

\subsection{Temperature of the environment, $T_{\mathrm{env}}$}

\begin{figure}
\centering
\includegraphics[width=8cm]{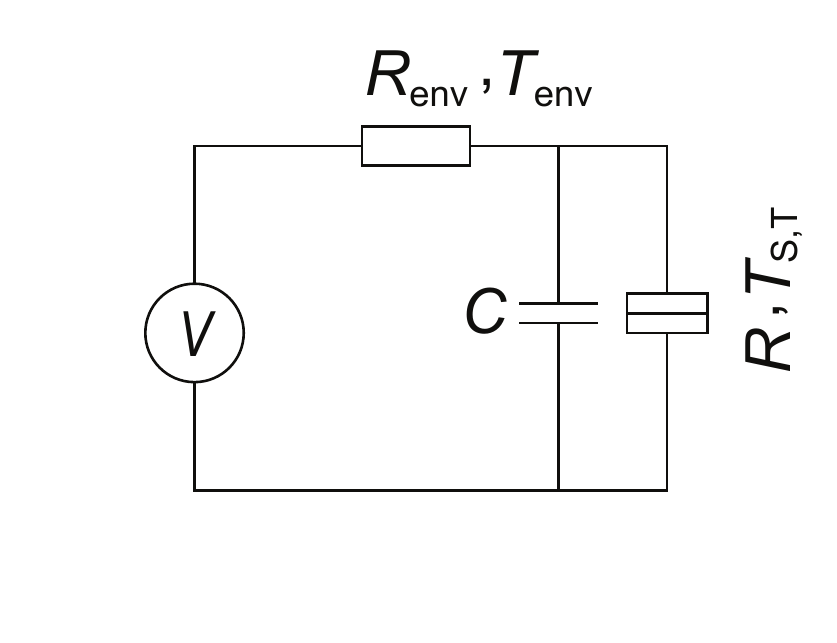}
\caption {Lumped element representation of the tunnelling junction and its environment. The tunnelling junction with tunnelling resistance $R$ and temperature $T$ is shunted by the junction capacitance $C$ and connected to a purely dissipative impedance $R_{\mathrm{env}}$ whose temperature is $T_{\mathrm{env}}$.}
\label{lump} 
\end{figure}

According to $P(E)$ theory, the environment of the STM junction is essentially the source of the fluctuations of the junction's phase $\tilde{\phi}$ (see Eq. \ref{PHI} in the Methods section), which couple to the charge degree of freedom $Q$ of the junction and thus affect the tunnelling process. The use of the fluctuation-dissipation theorem enables one to represent the fluctuating environment by a dissipative impedance $R_{\mathrm{env}}$ connected in series to the tunnelling junction with resistance $R$ and capacitance $C$ as shown in the lumped element model \cite{Grabert1992} in Fig. \ref{lump}. Because the spectrum of fluctuations depends on the temperature $T_{\mathrm{env}}$ of the environment (see Eq. \ref{J0}), an analysis of experimental tunnelling spectra yields the value of $T_{\mathrm{env}}$ \cite{Martinis1993}. As was mentioned in the introduction, Josephson tunnelling spectroscopy provides the most direct access to the $P(E)$ function of the environment and hence $T_{\mathrm{env}}$. 

We exploited the unique capability of our instrument to perform temperature-dependent STS and recorded 91 differential conductance $dI/dV(V)$ spectra of the STM junction in the Josephson regime while increasing the temperature $T_{\mathrm{ADR}}$ of the  adiabatic demagnetization refrigerator from 30 to $600 \ \mathrm{mK}$. The spectra were acquired automatically with a rate of one spectrum per every 30 min. The measurement was done on an atomically clean superconducting Al(100) surface using a superconducting tip. The latter was prepared by gentle indentations of the PtIr tip (see Fig. \ref{tip}) into the Al(100) surface. Visual inspection of the spectra in Fig. \ref{joseph} reveals that the Josephson conductance peak, located at zero bias \cite{Grabert1992} and visible in all spectra up to $T_{\mathrm{ADR}} \approx$ 500 mK, shows no noticeable thermal dependence, except for its eventual disappearance above 500 mK, which most likely is a natural consequence of the loss of superconductivity in the tip. This finding is surprising, considering that $T_{\mathrm{ADR}}$ experiences an almost 20-fold increase.

In an attempt to rationalize the absence of the thermal dependence in the data, we recall that $P(E)$, which according to Eq. \ref{I_JO} defines the $I(V)$ and hence the $dI/dV(V)$ spectra of the Josephson tunnelling junction, depends on $T_{\mathrm{env}}$ rather than $T_{\mathrm{T,S}}$ \cite{Martinis1993, Siewert1997}. Thus, the absent temperature broadening indicates that $T_{\mathrm{env}}$ is decoupled from $T_{\mathrm{T,S}}$. Indeed a situation were $T_{\mathrm{env}} \gg T_{\mathrm{T,S}}$ could naturally be realized due to an insufficient radiation shielding of the junction from the higher temperature stages of the cryostat \cite{Hergenrother1994, Hergenrother1995}. 

\begin{figure}
\centering
\includegraphics[width=8cm]{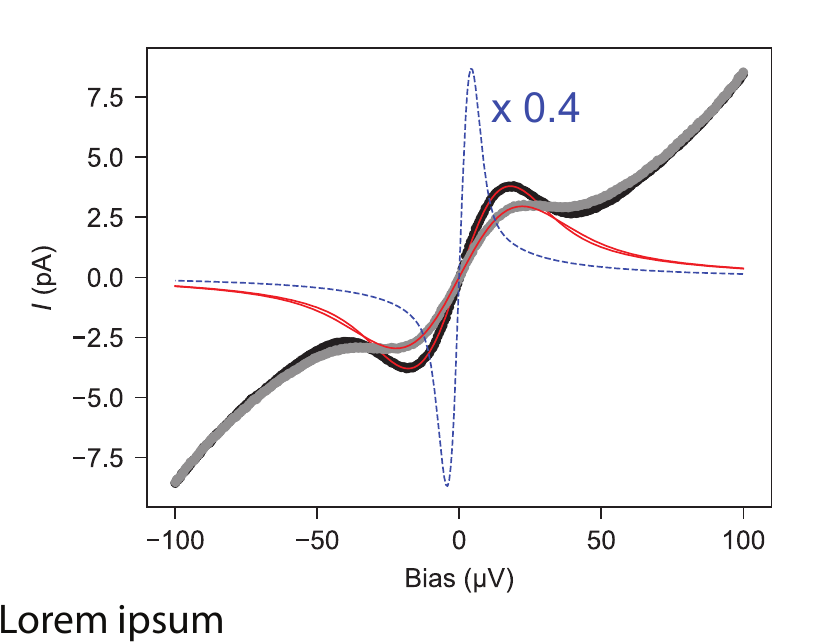}
\caption {The black data series depicts the Josephson $I(V)$ spectrum acquired simultaneously with the lowest temperature $dI/dV$ curve from Fig. \ref{joseph} measured at 34 mK and fitted with $T_{\mathrm{env}}=1.5$ K (red curve cf. text). Measured by the positions of the $I(V)$ extrema in this curve, the broadening due to environmental effects is $35$ $\mu$V. The blue dashed line shows the $I(V)$ spectrum of a junction in a well-thermalized environment obtained with the same parameters as for the black curve (cf. text) but $T_{\mathrm{env}}=0.1$ K. The corresponding value of environmental broadening is $9\ \mu$V. Gray data correspond to the $I(V)$ spectrum of the Josephson junction measured at 34 mK with the high-frequency filter of the bias line \cite{Esat2021} unplugged. The data were fitted with $T_{\mathrm{env}}=2.3$ K, keeping all other parameters unchanged (cf. text). Measured by the positions of the $I(V)$ extrema in this curve, the broadening due to the environmental effects is $44$ $\mu$V.}
\label{Joseph_fit} 
\end{figure} 

To test the plausibility of the $T_{\mathrm{env}} \gg T_{\mathrm{T,S}}$ scenario, we used the $P(E)$ theory to fit the experimental $I(V)$ curve acquired simultaneously with the lowest temperature $dI/dV(V)$ spectrum from  Fig. \ref{joseph} recorded at $T_{\mathrm{ADR}}=34$ mK. As Fig. \ref{Joseph_fit} shows, presumably due to the finite size effects in our superconducting tip, we registered a considerable quasiparticle tunnelling background in the $I(V)$ data. To keep the number of fitting parameters small, we simply neglected the presence of the quasiparticle background, by constraining the fitting to the [-25,25] $\mu$V interval of the bias voltage.

According to the Eqs. \ref{I_JO}, \ref{P_E}, \ref{J0}-\ref{ZT}, in the simplest case where the effective impedance of the circuit can be set to a real value $R_{\mathrm{env}}$, the simulation of the experimental $I(V)$ data from Fig. \ref{Joseph_fit} needs four fit parameters: $C$, $R_{\mathrm{env}}$, $E_{\mathrm{J}}$ and $T_{\mathrm{env}}$. Making an unconstrained fit with these parameters is, however, impossible as $C$ and $T_{\mathrm{env}}$ are coupled to each other, as can be spotted in Eq. \ref{PN}d defined in terms of $T/C$ ratio. Therefore, we fix $C=100$ fF in accordance with the BEM simulations of the STM junction discussed above, while freely varying the remaining three parameters $R_{\mathrm{env}}$, $E_{\mathrm{J}}$ and $T_{\mathrm{env}}$. Note that, in principle, $E_{\mathrm{J}}$ could be obtained from the experimental data using the Ambegaokar-Baratoff formula \cite{Ambegaokar1963, Jack2016}. However, here we used $E_{\mathrm{J}}$ as a fit parameter, because our data did not yield a reliable estimate of the superconducting gap $\Delta_{\mathrm{T}}$ in the tip, necessary for the evaluation of $E_{\mathrm{J}}$.

\begin{figure}
\centering
\includegraphics[width=8cm]{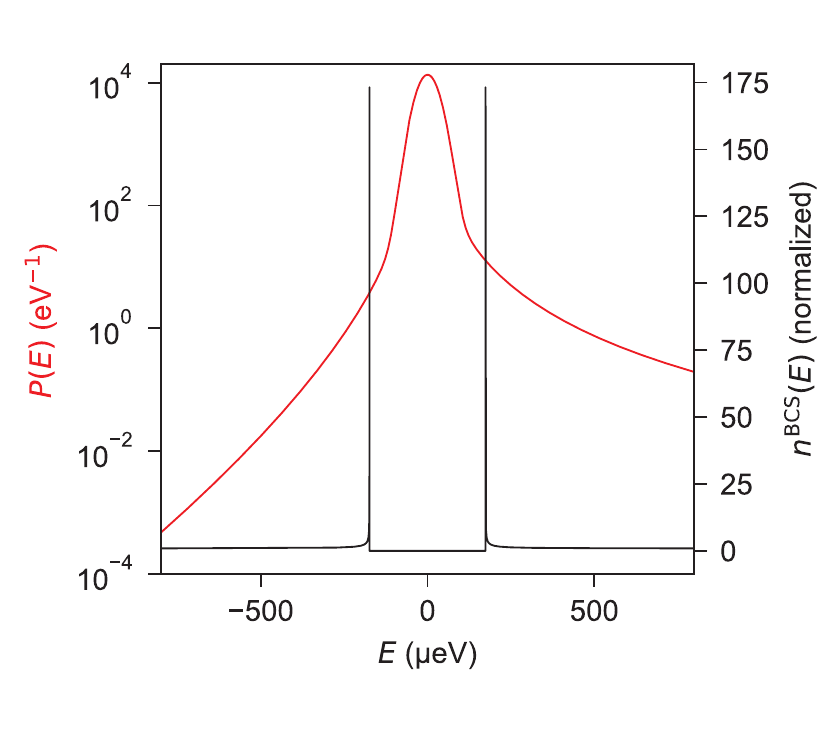}
\caption{The $P(E)$ function (red line) calculated for $T_{\mathrm{env}}=1.5$ K, $R_{\mathrm{env}}=31$ $\Omega$, $E_{\mathrm{J}}=8 \ \mu$eV and $C=100$ fF. 
The black curve displays $n^{\mathrm{BCS}}(E)$ for $\Delta=173 \ \mu$eV. The height of the quasiparticle peaks is limited by the finite size of the energy scale used in the calculation.
}
\label{pat} 
\end{figure}

Fig. \ref{Joseph_fit} exhibits the experimental $I(V)$ spectrum of the Josephson junction measured at 34 mK (black curve). The curve clearly demonstrates the signature of the Josephson super current in the presence of thermal fluctuations in the environment \cite{Joyez1999}. We fit the experimental curve by calculating the $P(E)$ function according to the numerical procedure proposed by Ingold and Grabert \cite{Ingold1991}  and also used by Ast and coworkers \cite{Ast2016}. The fit (red curve) yields $T_{\mathrm{env}}=1.5$ K, $R_{\mathrm{env}}=31$ $\Omega$ and $E_{\mathrm{J}}=8 \ \mu$eV. The $P(E)$ function corresponding to the fit is shown in Fig. \ref{pat}. The value of $T_{\mathrm{env}}$ obtained from the fit indicates that the scenario in which a cold mK STM junction finds itself inside a much warmer environment is indeed feasible. One could even speculate that the obtained value of $T_{\mathrm{env}}$ indicates the presence of thermal radiation from the 1 K stage of the cryostat. To visualize the effect that the environment has on the tunnelling, we plotted in Fig. \ref{Joseph_fit} another $I(V)$ (blue dashed) curve simulated with $T_{\mathrm{env}}=0.1$ K, i.e. for the case of a well-shielded junction with $T_{\mathrm{env}} \approx T_{\mathrm{T,S}}$. Fig. \ref{Joseph_fit} shows that a better shielding of the junction from the hot radiation could improve the spectroscopic resolution considerably.

The thermal radiation arriving from the hotter stages of the cryostat affects the tunnelling by coupling to the junction via its electrodes acting like an antenna \cite{Ast2016}. However, the high-frequency noise from the room temperature electronics should also couple to the junction, producing an effective increase of $T_{\mathrm{env}}$. We exploited a convenient possibility to demonstrate this effect experimentally and re-measured the $I(V)$ spectrum of the Josephson junction at $T_{\mathrm{ADR}}=34$ mK with the Pi-filter of the bias line (see Methods section) unplugged. The spectrum measured without the filter shown as the grey curve in Fig. \ref{Joseph_fit} was fitted with $T_{\mathrm{env}} = 2.3$ K, keeping all other parameters fixed. Thus, unplugging the filter changes $T_{\mathrm{env}}$ and worsens the experimental resolution from 35 $\mu$V to 44 $\mu$V, as measured by the extrema of the Josephson current feature in Fig. \ref{Joseph_fit}.

\subsection{Temperature of the tip}

Having determined the influence of the junction environment, we now turn to the problem of estimating $T_{\mathrm{T, S}}$, which, as was mentioned above, are the temperatures appearing in the Fermi-Dirac distributions of the tip and the sample electrodes. Generally speaking, the tip and the sample can have different temperatures, i.e. $T_{\mathrm{T}} \neq T_{\mathrm{S}}$. Therefore, to fully characterize the junction temperature, it could be necessary to perform two different experiments. As mentioned above, $T_{\mathrm{S}}$ can be obtained by STS with a superconducting tip on a normal metal sample, while $T_{\mathrm{T}}$  follows from STS with a normal metal tip on a superconducting sample. Assuming that the tip should have a higher temperature in our STM design, we only evaluate $T_{\mathrm{T}}$ as the upper bound on the junction temperature. 

We remark that despite a consensus about the utility of $n^{\mathrm{BCS}}(E)$ for estimating the junction temperature, there is no general agreement on the details of the analysis, with most variations occurring at the stage of fitting the experimental data. The problem is that using Eqs.\ref{BCS} and \ref{G_NIS} does not usually produce satisfactory fits, and as a consequence one has to employ expressions for the superconducting DOS that include additional parameters. The most commonly used expression of that type \cite{Song2010,Assig2013,Roychowdhury2014,Machida2018,Balashov2018,VonAllworden2018,Wong2020,Esat2021} was introduced by Dynes et al. \cite{Dynes1978} to account for the finite quasiparticle lifetime in a superconductor. It follows from Eq. \ref{BCS} by introducing a phenomenological parameter $\gamma$

\begin{equation} 
n_{\mathrm{S}}^{\mathrm{D}}=\mathrm{Re}\bigg[\frac{E-i\gamma}{\sqrt{(E-i\gamma)^2-\Delta^2}}\bigg].
\label{Dynes}
\end{equation}

 \noindent Although $n_{\mathrm{S}}^D$ indeed results in better fits of the STS data, the physical significance of $\gamma$ remains unclear, which makes the fitting procedure ambiguous.

\begin{figure}
\centering
\includegraphics[width=8cm]{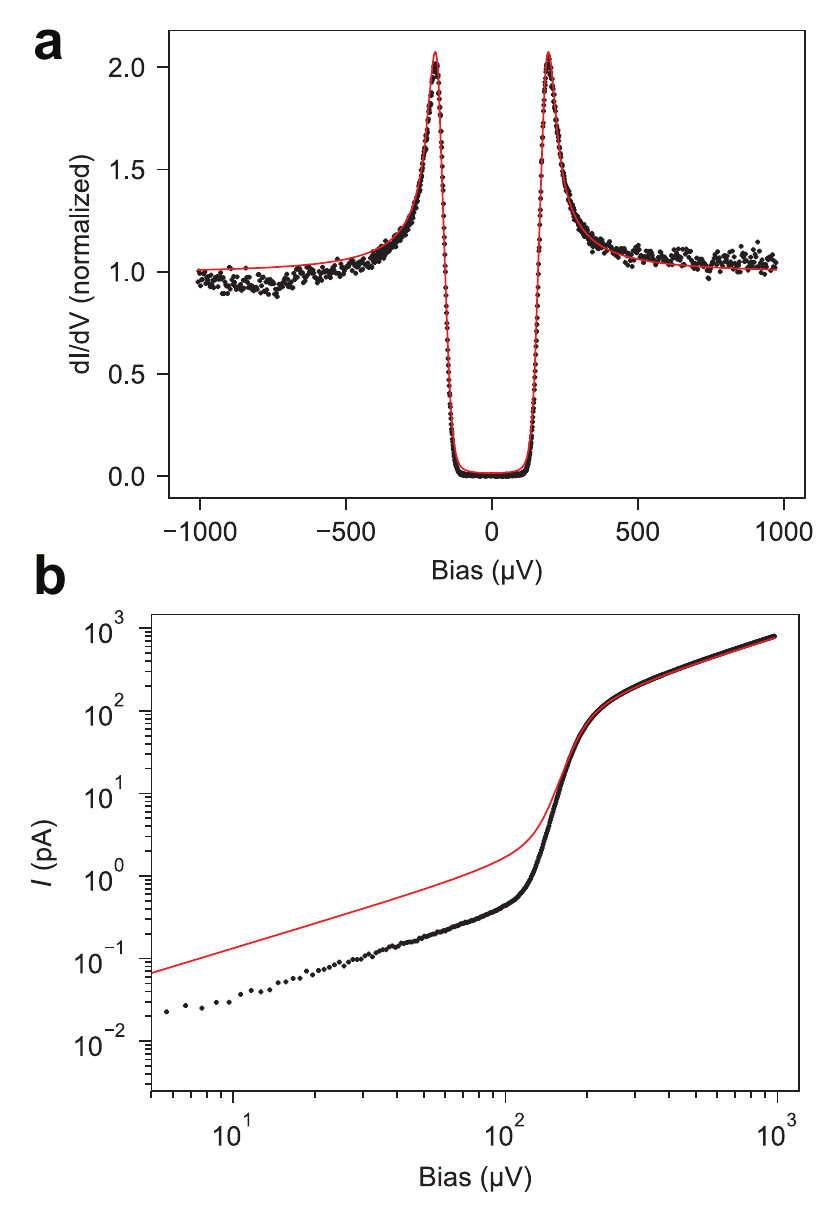}
\caption{(\textbf{a}) Raw $dI/dV(V)$ spectrum of the superconducting Al(100) surface measured with a normal-metal tip at 44 mK and fitted with Eq. \ref{G_NIS}, where $n_{\mathrm{S}}^{\mathrm{BCS}}$ was substituted by the Dynes expression for $n_{\mathrm{S}}^{\mathrm{D}}$ defined in Eq. \ref{Dynes}. For values of the fit parameters cf. text. (\textbf{b}) The $I(V)$ curve measured simultaneously with the $dI/dV(V)$ spectrum of panel (a) and fitted using the same parameter set used for calculating the fit in (a) complemented by the high-voltage or normal-state resistance of the junction $R=1.2$ M$\Omega$.
}
\label{Bad_fits} 
\end{figure} 

Besides the Dynes expression in Eq. \ref{Dynes}, there have been attempts to fit the experimental STS of superconducting gaps with the so-called Maki formula, which considers the effects of magnetic scattering \cite{Assig2013, Machida2018}. Inspection of the original work of Maki  \cite{Maki1964} reveals, however, that their approach applies to dirty superconductors, for which the mean free path of an electron is smaller than the coherence length of a Cooper pair. Because our clean single-crystal Al(100) sample does not fulfill this condition, we only analyze our data using the Dynes expression in Eq. \ref{Dynes}.

In Fig. \ref{Bad_fits}a we present a $dI/dV$ spectrum of the superconducting gap measured on Al(100) with a normal-metal PtIr tip at $T_{\mathrm{ADR}}=$ 44 mK. As expected for this temperature, the curve features a fully developed superconducting gap and two sharp conductance spikes situated symmetrically around zero bias. First, following the standard approach we fit the spectrum with Eq. \ref{G_NIS} in which we substitute $n_{\mathrm{BCS}}$ with $n_{\mathrm{S}}^{\mathrm{D}}$ and use $\Delta$ and $\gamma$ (see Eq. \ref{Dynes}) as two independent fit parameters, while the Fermi-Dirac distribution in Eq. \ref{G_NIS} contributes $T_{\mathrm{T}}$ as the third fit parameter. The $P(E)$ function in Eq. \ref{G_NIS} introduces no additional fit parameters, as it is calculated with the fixed set of parameters determined earlier from the Josephson junction data analysis. Note that in using Eq. \ref{G_NIS} we neglect the small effect of the 4 $\mu$V lock-in bias modulation (see Methods). 

\begin{figure}
\centering
\includegraphics[width=8cm]{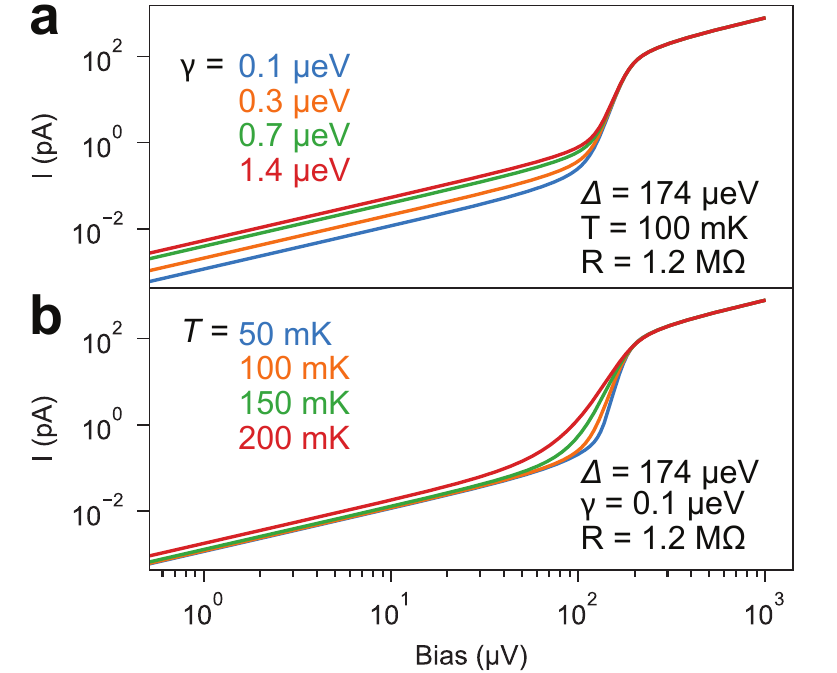}
\caption{Illustration of how the $I(V)$ function simulated using the Dynes expression in Eq. \ref{Dynes} reacts to changes in $T$ and $\gamma$. The respective parameters are listed in the insets.}
\label{Dynes_work} 
\end{figure} 

Fig. \ref{Bad_fits}a shows the fit of the superconducting gap obtained with $\Delta=175 \ \mu$eV, $\gamma=2.4 \ \mu$eV,  and $T_{\mathrm{T}}=99$ mK. Notice that the value of $T_{\mathrm{T}}$ comes out substantially lower than the one we reported earlier \cite{Esat2021}, which occurs mostly due to the inclusion of the $P(E)$ broadening effect into the consideration. The conventional fitting approach stops here, without giving detailed attention to the value of $\gamma=2.4 \ \mu$eV, which seems to be an order of magnitude too high, if compared, e.g., to the data reported for mesoscopic junctions \cite{Pekola2010}. 

However, the unusually high $\gamma$ causes a problem manifesting itself as a high in-gap conductance visible by closer inspection of the fit in Fig. \ref{Bad_fits}a. Technically, the problem arises because the least-square fitting routine minimizes the total sum of \textit{absolute} quadratic deviations $\sum_i \sqrt{x^2_i-y_i^2}$ between the data $x_i$ and the fit $y_i$ calculated at every experimental point $i$. Because the minimized deviation is \textit{absolute}, the points where the $x_i$ are large contribute more to its sum. Conversely, the points inside the gap, where the $x_i$ are small, produce a smaller contribution; hence, the deviations between the fit and the data inside the gap are less significant for minimizing that sum. 

To exhibit the extent of the problem clearly, we show in Fig. \ref{Bad_fits}b a log-log plot of the $I(V)$ curve measured simultaneously with the $dI/dV$ data in Fig. \ref{Bad_fits}a. The latter was fitted with a curve that has been generated with the same parameters as in Fig. \ref{Bad_fits}a, but complemented with the high-voltage (i.e. normal-state) resistance $R$ = 1.2 $\mathrm{M\Omega}$ of the junction. Fig. \ref{Bad_fits}b confirms that the fitting of the experimental data inside the gap needs improvement. 

\begin{figure}
\centering
\includegraphics[width=8cm]{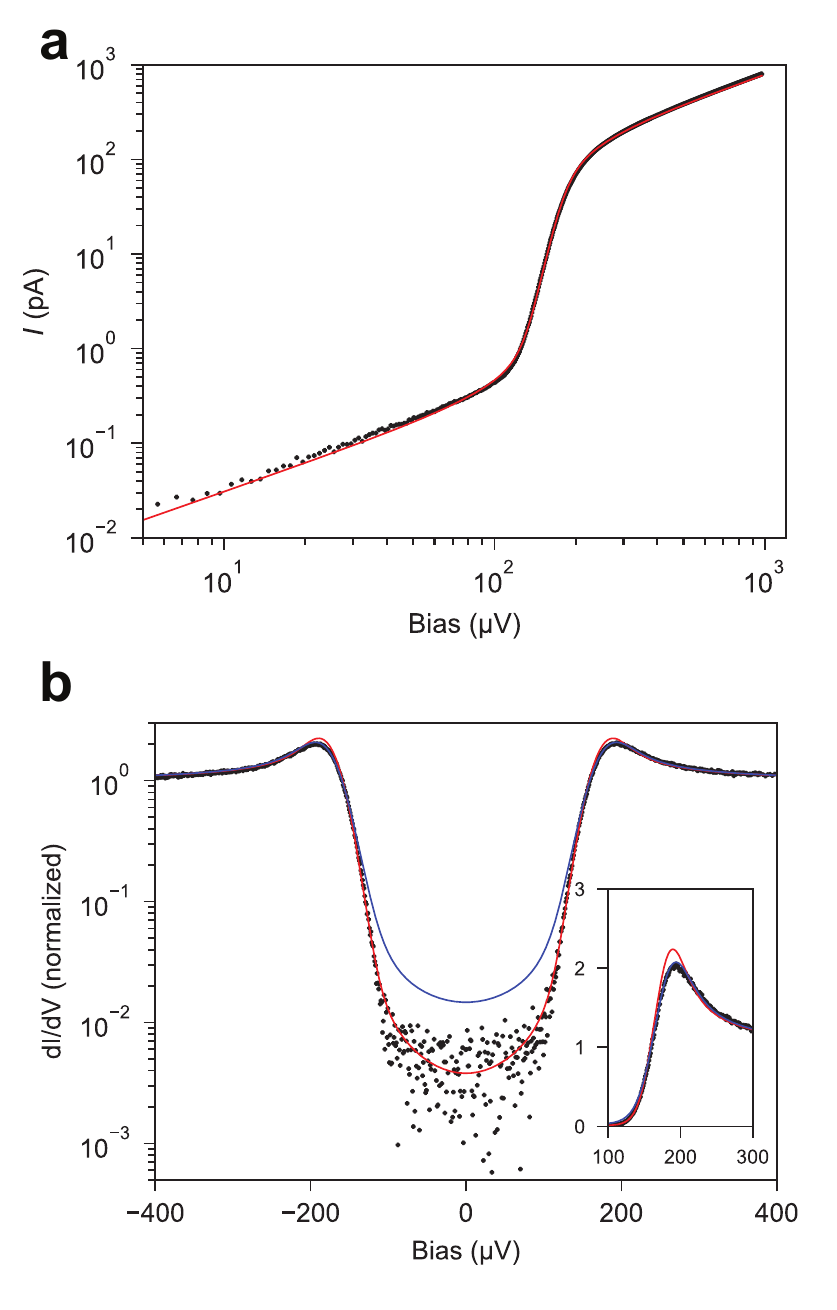}
\caption{(\textbf{a}) Fit of the $I(V)$ curve from Fig. \ref{Bad_fits}b, achieved with the modified fitting routine (cf. text). (\textbf{b}) $dI/dV(V)$ spectrum corresponding to the $I(V)$ curve of panel (a) plotted on a logarithmic scale to show the quality of the fit. The red curve is generated with the same parameter set as in panel (a). The blue curve is the fit from Fig. \ref{Bad_fits}a. (inset) Quasiparticle peak feature plotted on a linear scale.}
\label{Good_fit} 
\end{figure} 

Before we fix the problem with the fitting, it is instructive to inspect how the Dynes-generated $I(V)$ curve responds to changes in the parameters $\gamma$ and $T$. Fig. \ref{Dynes_work}a shows that for small $T$ and $\gamma$, the value of $\gamma$ defines primarily the zero-bias conductance $G_{V=0}$ of the junction. $T$, on the other hand, as Fig. \ref{Dynes_work}b shows, controls the sharpness of the transition between the low and high conductance regimes. Interestingly, as $T$ increases, it smears the gap edge towards zero bias and thus also raises the zero-bias conductance of the junction. Comparing Fig. \ref{Bad_fits}b and Fig. \ref{Dynes_work}a-b, we see that the value of $\gamma$ that we obtained from the first fitting attempt is indeed too high.

To fit the in-gap data correctly, we first change to fitting $I(V)$ instead of $dI/dV(V)$ curves. Besides that, we minimize the sum of \textit{relative} quadratic deviations $\sum_i \sqrt{\big(\frac{x_i-y_i}{x_i+y_i}\big)^2}$. The modified fit obtained with $\Delta=173 \ \mu$eV, $\gamma=0.5 \ \mu$eV, $T_\mathrm{T} = 78$ mK and $R_\mathrm{T}= 1.2 \ \mathrm{M\Omega}$ in Fig. \ref{Good_fit}a exhibits a much better agreement in the gap. Nevertheless, the improvement in the gap comes at the expense of the fit quality at the quasiparticle peaks, where, as the inset of Fig. \ref{Good_fit}b shows, the new fit overshoots the experiment appreciably. 

Concluding that the Dynes-generated curves cannot simultaneously fit all the features of the experimental superconducting gap spectrum, we come to the question of which part is more relevant for our analysis. We propose that the in-gap part of the spectrum is of greater importance because the quasiparticle peaks, as was also recently suggested by  Schwenk et al.\cite{Schwenk2020}, may experience an additional non-intrinsic energy-dependent broadening. One could further speculate that the observed broadening may stem from non-equilibrium effects in the junction, e.g., local heating of the tip by the current of tunnelling electrons \cite{Nahum1994}. If this indeed was the case, the temperature increase due to such heating could be about 20 mK, as Fig. \ref{Good_fit}b suggests.

\begin{figure*}
\centering
\includegraphics[width=16cm]{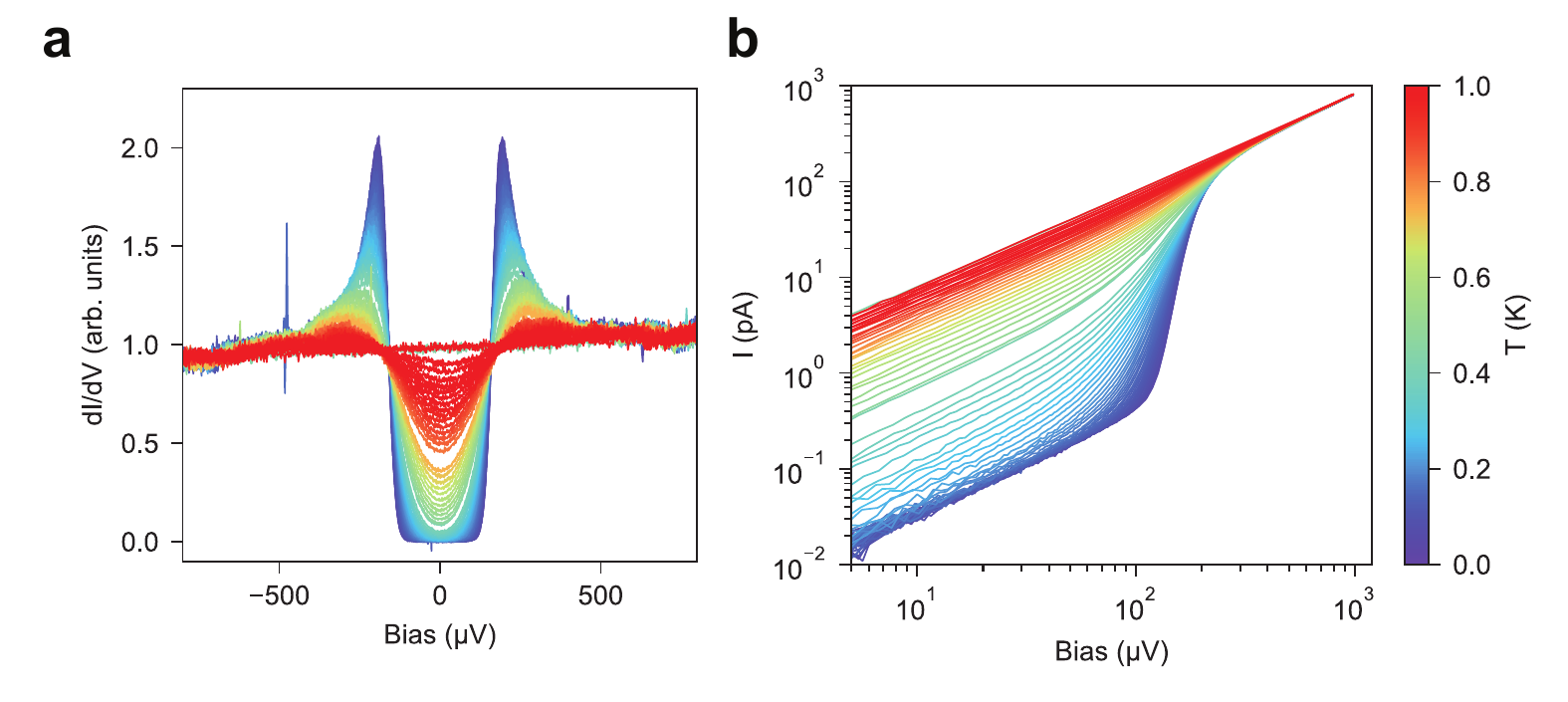}
\caption{The experimentally measured temperature dependence of $dI/dV(V)$ (\textbf{a}) and $I(V)$ (\textbf{b}) of the normal metal - superconductor tunnelling junction. The data were measured in a fully automatic mode, taking a spectrum every 30 minutes during ca. 48 hours. The noise visible in the $dI/dV(V)$ data comes from the few curves that were affected by small instabilities of the junction, caused by rapid pressure variations in the He exhaust line.
}
\label{T_series} 
\end{figure*} 

\begin{figure}
\centering
\includegraphics[width=8cm]{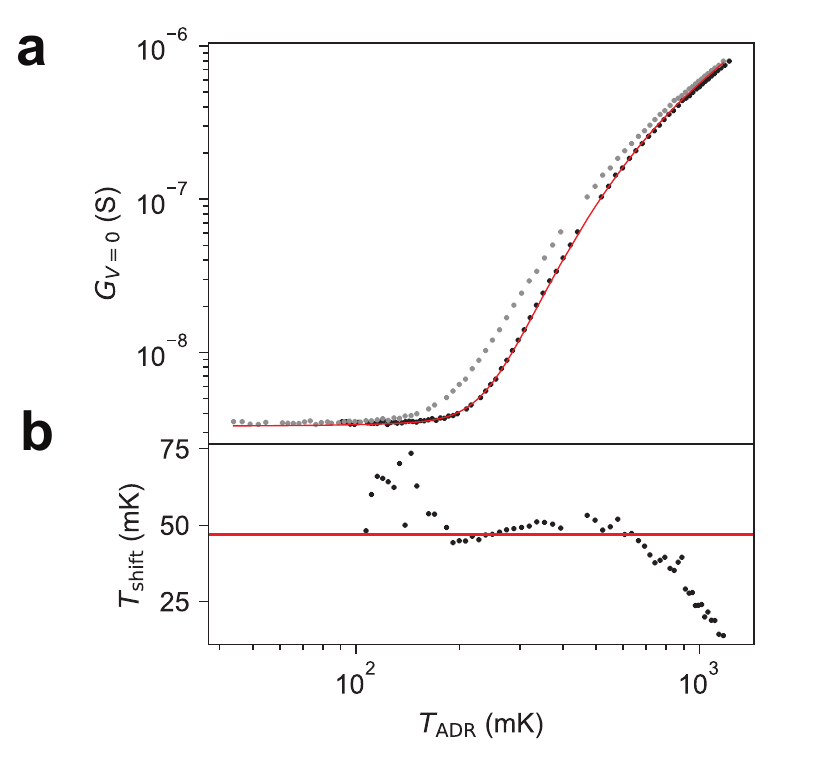}
\caption{(\textbf{a}) Thermal dependence of the zero-bias conductance $G_{V=0}$ obtained from the experiment (grey dots) and simulated (red line). The black dots belong to the same experiment as the gray ones, but are shifted by 45 mK to higher temperatures. (\textbf{b}) A shift between $T_{\mathrm{ADR}}$ and the temperature of the tip $T_{\mathrm{T}}$ was obtained from the data in (a). Red horizontal line marks the value $T_{\mathrm{shift}}=45$ mK.
}
\label{T_scale} 
\end{figure}

Developing the argument further and recalling Fig. \ref{Dynes_work}b, which shows how the zero-bias conductance $G_{V=0}(T)$ reacts to the increasing temperature, we wonder whether $G_{V=0}(T)$ alone could provide a good estimate of the temperature in the tip. To explore this question, we measured the junction characteristics, varying $T_{\mathrm{ADR}}$ between 44 mK and 1.2 K. As expected, the data plotted in Fig. \ref{T_series}a-b clearly show a gradual disappearance of the superconducting gap upon increasing $T_{\mathrm{ADR}}$. The experimental $G_{V=0}(T_{\mathrm{ADR}})$ plot in Fig. \ref{T_scale}a extracted from the data in Fig. \ref{T_series} reveals two regimes: Below $T_{\mathrm{ADR}} \sim 200$ mK $G_{V=0}$ stays almost constant, with its value determined by $\gamma$. At higher $T_{\mathrm{ADR}}$, $G_{V=0}$ increases quickly, indicating that the smearing of the gap reaches zero bias, similar to Fig. \ref{Dynes_work}b.

To simulate the $G_{V=0}(T_{\mathrm{ADR}})$ curve with Eq. \ref{Dynes}, we assumed that $\gamma$ is independent of the temperature and used the value $\gamma=0.56 \ \mu$eV. Second, we calculated $\Delta(T)$ according to a well-known analytic formula \cite{Gasparovic1966} by taking the critical temperature $T_\mathrm{c}$ of Al to be 1.2 K and using $\Delta(T=0)$ = 173 $\mu$eV as obtained from the fit in Fig. \ref{Good_fit}a. Plotting the calculated $G_{V=0}(T)$ curve in Fig. \ref{T_scale}a next to the experiment, we find a striking agreement between both curves if the experimental temperature scale  $T_{\mathrm{ADR}}$  is shifted by $T_\mathrm{shift}$ = 45 mK towards higher temperatures.  

The data from Fig. \ref{T_scale}a can be also plotted to show a $T_\mathrm{shift}$ for each individual experimental data point, which results in the $T_\mathrm{shift}(T_\mathrm{ADR})$ curve plotted in Fig. \ref{T_scale}b. Apparently, our $G_{V=0}(T)$ analysis yields a reasonably precise estimate of $T_\mathrm{shift}$ in the range where $T_\mathrm{ADR} \gtrsim 200$ mK. Due to the flattening of the $G_{V=0}(T)$ curve, however, the precision for $T_\mathrm{ADR} < 200$ mK deteriorates, but it seems that one could nevertheless safely assume that $T_\mathrm{shift}$ remains roughly constant towards lower-temperatures. Interestingly, as $T_\mathrm{ADR}$ increases, $T_\mathrm{shift}$ tends to diminish, which could happen due to increasing heat conductivity of the materials, leading to a better thermalization of the tip.

Looking back at Fig. \ref{Good_fit} that displays the $I(V)$ curve measured at $T_\mathrm{ADR}=44$ mK and recalling that $T_\mathrm{T}=78$ mK extracted from its fit, we can now compare this latter temperature to the $T_\mathrm{T}=T_\mathrm{ADR}+T_\mathrm{shift}=44 \ \mathrm{mK} +45 \ \mathrm{mK} = 89$ mK predicted by the analysis of $G_{V=0}(T)$.  Pondering the discrepancy between the predicted values, we note that due to the mentioned inaccuracy of the $G_{V=0}(T)$ approach below 200 mK, the result from the fit of the full $I(V)$ curve should probably be given more credence. Note, however, that unlike the $G_{V=0}(T)$ analysis, the full $I(V)$ curve fitting will generally work much less reliably at higher temperatures due to the washing out of the gap.  

Finally, we conclude our analysis by discussing $\gamma$, the microscopic origin of which, especially for temperatures far below $T_\mathrm{c}$, is not yet well-understood. Recently, Pekkola et al. \cite{Pekola2010} proposed that $\gamma$ may reflect the strength of photon-assisted tunnelling that occurs at energies far below the gap edge. Assuming that our environment has a blackbody spectrum with $T_{\mathrm{env}}=1.5$ K, we obtain $2.7 \times T_{\mathrm{env}} = 4.25$ K as the average energy of a photon. This is substantially higher than $\Delta/k_{\mathrm{B}} \approx 2 \ \mathrm{K}$, which points towards the possibility of photon assisted tunnelling. 

According to Eqs.\ref{Dynes} and \ref{G_NIS} the dimensionless parameter $\gamma/\Delta$ is directly related to the ratio of the junction conductance at zero bias to its conductance at high bias, $\gamma/\Delta \simeq R/R_0$. Due to the presence of the superconducting gap, tunnelling at zero bias can only occur if an electron absorbs energy larger than $\Delta$ from the environment. Because the probability of such an event is defined by the $P(E)$ function for $E \leq -\Delta$, Di Marco et al. \cite{DiMarco2013} proposed to evaluate $\gamma/\Delta$ as

\begin{equation} 
\gamma/\Delta \simeq 2\int^{-\Delta}_{-\infty}dE \ n^{\mathrm{BCS}}(E)P(E),
\label{int_pat}
\end{equation}

Using the parameters determined previously (by fitting to experimental data,  see above), the two functions under the integral in Eq. \ref{int_pat} are plotted in Fig. \ref{pat}. Evaluating the integral numerically yields $\gamma/\Delta = 4 \times 10^{-4}$, while the experimental fit in Fig. \ref{Good_fit} provides $\gamma/\Delta = 3\times 10^{-3}$. We regard this discrepancy as rather small, taking the phenomenological character of the procedure for evaluating the $P(E)$ function into account. Thus, the data favour the idea of Pekola et al. \cite{Pekola2010} that $\gamma$ of the Dynes fit reflects the strength of the photon-assisted tunnelling. Note that this effect could naturally occur in our STM junction, because it is not shielded from the radiation of the 1 K stage of the cryostat \cite{Esat2021}.

In conclusion, our temperature-dependent STS data acquired with the Al/vacuum/Al(100) Josephson junction between 30 and 600 mK demonstrated no appreciable thermal broadening effects. We propose that this occurs due to the difference between the temperature $T_{\mathrm{env}}$ of the environment and the junction temperature $T_{\mathrm{T, S}}$ with $T_{\mathrm{env}} \gg T_{\mathrm{T, S}}$. Fitting the Josephson STS data, we obtained the $P(E)$ function that characterizes the probability of the energy exchange between the environment and the junction. According to the fit, the temperature of the junction's environment is $T_{\mathrm{env}}=1.5$ K, substantially higher than the temperature of the ADR cryostat in which we performed the measurement. Although our data provide a qualitatively clear picture of a cold STM junction embedded in a hot environment, the obtained value of $T_{\mathrm{env}}$ needs to be treated with caution, due to the phenomenological character of the approach to calculating the $P(E)$ function and the crudeness with which we evaluated the junction capacitance $C$.

Using the tunnelling spectrum of the superconducting gap measured with a normal-metal tip on the Al(100) surface, we have demonstrated the potential fallibility of the usual $dI/dV(V)$ fitting, which does not pay sufficient attention to the low-conductance data inside the gap. We showed that minimizing the relative deviation helps achieving a much better fit of the in-gap data, albeit at the expense of the fit quality of the quasiparticle peaks. However, as we argued, the quasiparticle peaks may be subject to additional non-intrinsic and energy-dependent broadening mechanisms. Therefore, their accurate fitting may not be necessary for determining  the intrinsic parameters such as the junction's temperature and the Dynes $\gamma$. Using a modified fitting routine, we found that at $T_{\mathrm{ADR}} = 44$ mK the Fermi-Dirac distribution temperature in the tip was $T_{\mathrm{T}}=77$ mK and $\gamma=0.5 \ \mu$eV or $\gamma/ \Delta = 0.003$. 

We also proposed a new way of determining the temperature of the STM junction by analyzing the temperature scaling of the zero-bias conductance $G_{V=0}(T)$ only. Using this approach, we reconstructed $T_{\mathrm{T}}$ by mapping out its deviation $T_{\mathrm{shift}}$ from the cryostat temperature $T_{\mathrm{ADR}}$. Our data showed that $T_{\mathrm{shift}}\approx 45$ mK for $T<600$ mK. According to this method, the lowest $T_{\mathrm{T}}$ that we can currently achieve in our setup is $26 \ \mathrm{mK} + 45 \ \mathrm{mK} = 71 \ \mathrm{mK}$.

Finally, our data suggest that the value of the Dynes parameter $\gamma$ obtained from fitting the superconducting gap, in agreement with the proposal of Pekkola et al. \cite{Pekola2010, DiMarco2013}, reflects the strength of the photon-assisted tunnelling caused by the radiation arriving at the STM junction from the hotter stages of the cryostat. This finding emphasizes the importance of good radiation shielding and possibly low-temperature high-frequency filtering in the design of a mK STM. 

\section{Methods}

\subsection{Experimental setup}

We performed the experiments in a mK STM, the details and performance of which were reported recently \cite{Esat2021}. Thus, we only briefly list its essential characteristics here. The STM head is thermally well-anchored to a base plate of an adiabatic demagnetization refrigerator (ADR) that resides inside a UHV chamber. The ADR reaches a minimum temperature of 26 mK, measured by a calibrated RuOx sensor positioned on the base plate of the ADR right next to the STM head. We refer to the temperature reading of that thermometer as $T_{\mathrm{ADR}}$. ADR enables STM/STS experiments at varying $T_{\mathrm{ADR}}$ conditions spanning the range between 30 mK and 1-2 K. STM/STS can be performed in the presence of a magnetic field of up to $8 \ \mathrm{T}$ applied in the direction perpendicular to the sample surface. To reduce the high-frequency noise in the junction we use Pi-filters on the bias and all five high-voltage piezo lines. 

\subsection{STS}

The STS data discussed here were acquired on the clean surface of an Al(100) single crystal clamped to a custom-made flag-type sample holder. The sample holder body was fabricated from tungsten, while the clamping comprised a molybdenum foil and a set of molybdenum screws fixing the foil to the sample holder body. Due to the superconductivity of molybdenum, we made sure that the foil contained no closed holes to prevent flux trapping at low temperatures. The sample surface was prepared in UHV by repeated cycles of Ar$^+$ ion sputtering and subsequent annealing at 400$^\circ$C. We used low-energy electron diffraction (LEED) and Auger electron spectroscopy (AES) to verify the surface quality.

The Al(100) sample was superconducting during all of the STS measurements reported here. Interestingly, we could only observe the superconductivity by applying a compensating B field of 8-10 mT to the sample magnet. We speculate that the compensation field was necessary to neutralize the effect of the magnetic flux caught by one of the superconducting coils during the ADR run \cite{Esat2021}. 

The superconductivity in the sample always appeared abruptly upon ramping up the compensating B field. Moreover, the tunnelling spectrum of the superconducting gap did not change when the compensating sample B field was further varied. Instead, it disappeared and reappeared erratically. We rationalize this behavior by assuming that our sample was mainly in an intermediate state characterized by an ordered structure of macroscopically large superconducting domains separated by normal-metal areas \cite{Tinkham1996}. In such a state, slight variations of the sample B field can cause rearrangements of the domain structure, which, in turn, produce abrupt changes in the type of the domain (superconducting vs. normal) faced by the tip. 

The Josephson junction STS data were measured at 30 mK $\leq T_{\mathrm{ADR}} \leq 600$ mK with a compensating field of 8 mT kept throughout the measurement series. For the measurement, the PtIr tip was made superconducting by indenting it gently into the Al(100) surface until the Josephson conductance peak at zero bias (see Fig. \ref{joseph}) appeared in the conductance spectra. The series of spectra featuring the superconducting gap of the Al(100) surface (Fig. \ref{T_series}) was acquired with a clean PtIr tip at 40 mK $\leq T_{\mathrm{ADR}} \leq 1.2$ K. The lower temperature data were collected with the compensating field of 8 mT. At $T_{\mathrm{ADR}} \approx 460 \ \mathrm{mK}$, the sample superconductivity disappeared abruptly, and we regained it by tuning the compensating field to 10 mT, the value at which all higher temperature data were then collected.

STS was performed using the internal lock-in of the Nanonis software. The parameters of the lock-in were: modulation frequency 187.7 Hz, modulation amplitude 2 $\mu$V for the Josephson (Fig. \ref{joseph}) and 4 $\mu$V for the superconducting gap spectroscopy (Fig. \ref{T_series}). The tunnelling current was measured using a fixed gain $(10^{10})$ amplifier from NF Corporation \cite{NF}. All experimental data are displayed as raw data without any additional postprocessing. Each presented STS curve is a single sweep spectrum acquired in less than 220 s. 

\subsection{Fitting STS data with the $P(E)$ theory}

The $P(E)$ theory which was originally developed for mesoscopic tunnelling junctions \cite{Grabert1992}, derives the $P(E)$ function from the junction's equilibrium phase-phase correlation function $J(t) \equiv \langle [\tilde{\phi}(t)-\tilde{\phi}(0)]\tilde{\phi}(t) \rangle $,

\begin{equation} 
P(E)=\frac{1}{2 \pi \hbar} \int_{-\infty}^{\infty} dt \ \mathrm{exp} \bigg[J(t)+\frac{i}{\hbar}Et\bigg].
\label{P_E}
\end{equation}

According to the Wiener-Khinchin theorem (see Appendix of the ref. \cite{Clerk2010}), $P(E)$ is a noise power spectral density of the junction's charge shift operator $e^{-i\tilde{\phi}}$, where the phase $\tilde{\phi}(t)$ is defined as

\begin{equation} 
	\tilde{\phi}(t)=\frac{e}{\hbar} \left( \int_{-\infty}^tdt'U(t')-Vt \right),
\label{PHI}
\end{equation}

\noindent with $U=Q/C$ being the momentary voltage across the junction with capacitance $C$ charged by $Q$, and $V$ is the constant voltage applied by an external source considered to be ideal.

In their pioneering attempt to apply the $P(E)$ theory to the STM, Ast and coworkers extended the theory by pointing out that the fluctuations of the phase seen by the STM junction split into two contributions \cite{Ast2016}: $J(t)=J_0(t)+J_{\mathrm{N}}(t)$. According to the fluctuation-dissipation theorem, $J_0(t)$ is related to the imaginary part of the response function $\chi(\omega)=(e/ \hbar)^2 Z_t(\omega)/i\omega$, where $Z_t(\omega)$ is the total impedance of the circuit consisting of the the junction capacitance $C$ and the environmental impedance $Z(\omega)$ 
via \cite{Grabert1992}

\begin{equation}
\begin{split} 
J_0(t)=2\int^{\infty}_0 \frac{d\omega}{\omega}\frac{\mathrm{Re}[ Z_\mathrm{t}(\omega)]}{R_\mathrm{K}}\times\\
\times\bigg[\coth\bigg(\frac{1}{2}\frac{\hbar \omega}{k_\mathrm{B}T}\bigg)[\cos(\omega t)-1]-i\sin(\omega t) \bigg].
\end{split} 
\label{J0}
\end{equation}

\noindent The newly introduced term $J_\mathrm{N}(t)$ determines thermal voltage noise on the junction capacitor $C$. Splitting $J(t)$ into the two contributions helps express the $P(E)$ function as a convolution of two parts

\begin{equation} 
P(E)=\int dE' P_0(E-E')P_\mathrm{N}(E'),
\label{PO*PN}
\end{equation}

\noindent where $P_\mathrm{N}$ related to $J_{\mathrm{N}}(t)$ takes a simple Gaussian form

\begin{equation} 
P_\mathrm{N}(E)=\frac{1}{\sqrt{4\pi E_\mathrm{C}k_\mathrm{B}T}}\exp\bigg[ -\frac{E^2}{4E_\mathrm{C}k_\mathrm{B}T} \bigg],
\label{PN}
\end{equation}

\noindent with $E_\mathrm{C}=e^2/2C$ being the charging energy of the STM junction capacitor $C$. Note that for the Josephson current carried by Cooper pairs $E_\mathrm{C}=(2e)^2/2C$.

According to Eqs. \ref{J0} and \ref{PN} both $P_0(E)$ and $P_\mathrm{N}(E)$ depend on the temperature of the junction environment $T_\mathrm{env}$ \cite{Martinis1993}. Therefore, calculating $P(E)$ and using it in Eq. \ref{I_JO} for fitting the experimental $I(V)$ characteristics of a Josephson junction should yield $T_{\mathrm{env}}$.

While obtaining $P_\mathrm{N}(E)$ is straightforward, the calculation of $P_0(E)$, according to Eq. \ref{J0}, becomes a task of finding $Z_\mathrm{t}(\omega)$. Currently, there are no well-established schemes which allow doing this reliably for the case of an STM junction. Therefore, we follow the procedure developed for mesoscopic tunnelling junctions \cite{Grabert1992,Ast2016}. That approach starts by representing the junction and its environment as a lumped-element circuit with the tunnelling junction shunted by a capacitance $C$ and connected to an ideal voltage source via an environmental impedance $Z(\omega)$. Then the expression for $Z_\mathrm{t}(\omega)$ can be written as \cite{Grabert1992}

\begin{equation} 
Z_\mathrm{t}(\omega)=\frac{1}{i \omega C+1/Z(\omega)}.
\label{ZT}
\end{equation}

In the simplest case $Z(\omega)$ only has a dissipative part that is independent of $\omega$, i.e. $Z(\omega)=R_{\mathrm{env}}$ \cite{Grabert1992,Hergenrother1995,Pekola2010,DiMarco2013}. The value $R_{\mathrm{env}}$ determines the strength of coupling between the junction and its environment, with $R_{\mathrm{env}} \ll R_{\mathrm{Q}}\equiv h/e^2$ determining the weak coupling regime in which most tunnelling electrons leave the environmental modes, except those close to $\omega=0$,  undisturbed, the $P(E)$ function is thus peaked at $E=0$, and the charge transferred through the junction is removed almost instantaneously by the voltage source \cite{Grabert1992}. According to Eq. \ref{I_JO}, fitting of the $I(V)$ curves measured with a Josephson junction then needs four fit parameters: $C$, $R_{\mathrm{env}}$, $E_{\mathrm{J}}$ and $T_{\mathrm{env}}$.

\section{Acknowledgments}
We thank Christian Ast and Gianluigi Catelani for fruitful discussions. Funding: T.E., R.T. and F.S.T. acknowledge support from the German Federal Ministry of Education and Research through the funding program 13N16032. 

\section{Author contributions}
T.E., F.S.T., and R.T. conceived the experiment. T.E. and R.T. performed the measurements with the technical help from X.Y. and F.M.. T.E. and R.T. analyzed the experimental data. H.S. performed the BEM calculations of the STM junction capacitance. R.T. wrote the manuscript with significant input from F.S.T.. All authors discussed the results and reviewed the manuscript. Competing interests: The authors declare no competing interests. Data and materials availability: All data presented in this paper are publicly available through the J\"ulich Dataset Respository with the identifier xxx.


\section{References}


\begin{thebibliography}{47}%
\makeatletter
\providecommand \@ifxundefined [1]{%
 \@ifx{#1\undefined}
}%
\providecommand \@ifnum [1]{%
 \ifnum #1\expandafter \@firstoftwo
 \else \expandafter \@secondoftwo
 \fi
}%
\providecommand \@ifx [1]{%
 \ifx #1\expandafter \@firstoftwo
 \else \expandafter \@secondoftwo
 \fi
}%
\providecommand \natexlab [1]{#1}%
\providecommand \enquote  [1]{``#1''}%
\providecommand \bibnamefont  [1]{#1}%
\providecommand \bibfnamefont [1]{#1}%
\providecommand \citenamefont [1]{#1}%
\providecommand \href@noop [0]{\@secondoftwo}%
\providecommand \href [0]{\begingroup \@sanitize@url \@href}%
\providecommand \@href[1]{\@@startlink{#1}\@@href}%
\providecommand \@@href[1]{\endgroup#1\@@endlink}%
\providecommand \@sanitize@url [0]{\catcode `\\12\catcode `\$12\catcode
  `\&12\catcode `\#12\catcode `\^12\catcode `\_12\catcode `\%12\relax}%
\providecommand \@@startlink[1]{}%
\providecommand \@@endlink[0]{}%
\providecommand \url  [0]{\begingroup\@sanitize@url \@url }%
\providecommand \@url [1]{\endgroup\@href {#1}{\urlprefix }}%
\providecommand \urlprefix  [0]{URL }%
\providecommand \Eprint [0]{\href }%
\providecommand \doibase [0]{http://dx.doi.org/}%
\providecommand \selectlanguage [0]{\@gobble}%
\providecommand \bibinfo  [0]{\@secondoftwo}%
\providecommand \bibfield  [0]{\@secondoftwo}%
\providecommand \translation [1]{[#1]}%
\providecommand \BibitemOpen [0]{}%
\providecommand \bibitemStop [0]{}%
\providecommand \bibitemNoStop [0]{.\EOS\space}%
\providecommand \EOS [0]{\spacefactor3000\relax}%
\providecommand \BibitemShut  [1]{\csname bibitem#1\endcsname}%
\let\auto@bib@innerbib\@empty
%</preamble>
\bibitem [{\citenamefont {Wagner}\ and\ \citenamefont
  {Temirov}(2015)}]{Wagner2015}%
  \BibitemOpen
  \bibfield  {author} {\bibinfo {author} {\bibfnamefont {C.}~\bibnamefont
  {Wagner}}\ and\ \bibinfo {author} {\bibfnamefont {R.}~\bibnamefont
  {Temirov}},\ }\bibfield  {title} {\enquote {\bibinfo {title} {{Tunnelling
  junctions with additional degrees of freedom: An extended toolbox of scanning
  probe microscopy}},}\ }\href {\doibase 10.1016/j.progsurf.2015.01.001}
  {\bibfield  {journal} {\bibinfo  {journal} {Prog. Surf. Sci.}\ }\textbf
  {\bibinfo {volume} {90}},\ \bibinfo {pages} {194--222} (\bibinfo {year}
  {2015})}\BibitemShut {NoStop}%
\bibitem [{\citenamefont {Chen}, \citenamefont {Bae},\ and\ \citenamefont
  {Heinrich}(2022)}]{Chen2022}%
  \BibitemOpen
  \bibfield  {author} {\bibinfo {author} {\bibfnamefont {Y.}~\bibnamefont
  {Chen}}, \bibinfo {author} {\bibfnamefont {Y.}~\bibnamefont {Bae}}, \ and\
  \bibinfo {author} {\bibfnamefont {A.~J.}\ \bibnamefont {Heinrich}},\
  }\bibfield  {title} {\enquote {\bibinfo {title} {{Harnessing the Quantum
  Behavior of Spins on Surfaces}},}\ }\href {\doibase 10.1002/adma.202107534}
  {\bibfield  {journal} {\bibinfo  {journal} {Adv. Mater.}\ ,\ \bibinfo {pages}
  {2107534}} (\bibinfo {year} {2022})}\BibitemShut {NoStop}%
\bibitem [{\citenamefont {Song}\ \emph
  {et~al.}(2010{\natexlab{a}})\citenamefont {Song}, \citenamefont {Otte},
  \citenamefont {Shvarts}, \citenamefont {Zhao}, \citenamefont {Kuk},
  \citenamefont {Blankenship}, \citenamefont {Band}, \citenamefont {Hess},\
  and\ \citenamefont {Stroscio}}]{Song2010}%
  \BibitemOpen
  \bibfield  {author} {\bibinfo {author} {\bibfnamefont {Y.~J.}\ \bibnamefont
  {Song}}, \bibinfo {author} {\bibfnamefont {A.~F.}\ \bibnamefont {Otte}},
  \bibinfo {author} {\bibfnamefont {V.}~\bibnamefont {Shvarts}}, \bibinfo
  {author} {\bibfnamefont {Z.}~\bibnamefont {Zhao}}, \bibinfo {author}
  {\bibfnamefont {Y.}~\bibnamefont {Kuk}}, \bibinfo {author} {\bibfnamefont
  {S.~R.}\ \bibnamefont {Blankenship}}, \bibinfo {author} {\bibfnamefont
  {A.}~\bibnamefont {Band}}, \bibinfo {author} {\bibfnamefont {F.~M.}\
  \bibnamefont {Hess}}, \ and\ \bibinfo {author} {\bibfnamefont {J.~A.}\
  \bibnamefont {Stroscio}},\ }\bibfield  {title} {\enquote {\bibinfo {title}
  {Invited review article: A 10 {mK} scanning probe microscopy facility},}\
  }\href {\doibase 10.1063/1.3520482} {\bibfield  {journal} {\bibinfo
  {journal} {Rev. Sci. Instrum.}\ }\textbf {\bibinfo {volume} {81}},\ \bibinfo
  {pages} {121101} (\bibinfo {year} {2010}{\natexlab{a}})}\BibitemShut
  {NoStop}%
\bibitem [{\citenamefont {Misra}\ \emph {et~al.}(2013)\citenamefont {Misra},
  \citenamefont {Zhou}, \citenamefont {Drozdov}, \citenamefont {Seo},
  \citenamefont {Urban}, \citenamefont {Gyenis}, \citenamefont {Kingsley},
  \citenamefont {Jones},\ and\ \citenamefont {Yazdani}}]{Misra2013}%
  \BibitemOpen
  \bibfield  {author} {\bibinfo {author} {\bibfnamefont {S.}~\bibnamefont
  {Misra}}, \bibinfo {author} {\bibfnamefont {B.~B.}\ \bibnamefont {Zhou}},
  \bibinfo {author} {\bibfnamefont {I.~K.}\ \bibnamefont {Drozdov}}, \bibinfo
  {author} {\bibfnamefont {J.}~\bibnamefont {Seo}}, \bibinfo {author}
  {\bibfnamefont {L.}~\bibnamefont {Urban}}, \bibinfo {author} {\bibfnamefont
  {A.}~\bibnamefont {Gyenis}}, \bibinfo {author} {\bibfnamefont {S.~C.~J.}\
  \bibnamefont {Kingsley}}, \bibinfo {author} {\bibfnamefont {H.}~\bibnamefont
  {Jones}}, \ and\ \bibinfo {author} {\bibfnamefont {A.}~\bibnamefont
  {Yazdani}},\ }\bibfield  {title} {\enquote {\bibinfo {title} {Design and
  performance of an ultra-high vacuum scanning tunneling microscope operating
  at dilution refrigerator temperatures and high magnetic fields},}\ }\href
  {\doibase 10.1063/1.4822271} {\bibfield  {journal} {\bibinfo  {journal} {Rev.
  Sci. Instrum.}\ }\textbf {\bibinfo {volume} {84}},\ \bibinfo {pages} {103903}
  (\bibinfo {year} {2013})}\BibitemShut {NoStop}%
\bibitem [{\citenamefont {Assig}\ \emph {et~al.}(2013)\citenamefont {Assig},
  \citenamefont {Etzkorn}, \citenamefont {Enders}, \citenamefont {Stiepany},
  \citenamefont {Ast},\ and\ \citenamefont {Kern}}]{Assig2013}%
  \BibitemOpen
  \bibfield  {author} {\bibinfo {author} {\bibfnamefont {M.}~\bibnamefont
  {Assig}}, \bibinfo {author} {\bibfnamefont {M.}~\bibnamefont {Etzkorn}},
  \bibinfo {author} {\bibfnamefont {A.}~\bibnamefont {Enders}}, \bibinfo
  {author} {\bibfnamefont {W.}~\bibnamefont {Stiepany}}, \bibinfo {author}
  {\bibfnamefont {C.~R.}\ \bibnamefont {Ast}}, \ and\ \bibinfo {author}
  {\bibfnamefont {K.}~\bibnamefont {Kern}},\ }\bibfield  {title} {\enquote
  {\bibinfo {title} {A 10 {mK} scanning tunneling microscope operating in ultra
  high vacuum and high magnetic fields},}\ }\href {\doibase 10.1063/1.4793793}
  {\bibfield  {journal} {\bibinfo  {journal} {Rev. Sci. Instrum.}\ }\textbf
  {\bibinfo {volume} {84}},\ \bibinfo {pages} {033903} (\bibinfo {year}
  {2013})}\BibitemShut {NoStop}%
\bibitem [{\citenamefont {Roychowdhury}\ \emph {et~al.}(2014)\citenamefont
  {Roychowdhury}, \citenamefont {Gubrud}, \citenamefont {Dana}, \citenamefont
  {Anderson}, \citenamefont {Lobb}, \citenamefont {Wellstood},\ and\
  \citenamefont {Dreyer}}]{Roychowdhury2014}%
  \BibitemOpen
  \bibfield  {author} {\bibinfo {author} {\bibfnamefont {A.}~\bibnamefont
  {Roychowdhury}}, \bibinfo {author} {\bibfnamefont {M.~A.}\ \bibnamefont
  {Gubrud}}, \bibinfo {author} {\bibfnamefont {R.}~\bibnamefont {Dana}},
  \bibinfo {author} {\bibfnamefont {J.~R.}\ \bibnamefont {Anderson}}, \bibinfo
  {author} {\bibfnamefont {C.~J.}\ \bibnamefont {Lobb}}, \bibinfo {author}
  {\bibfnamefont {F.~C.}\ \bibnamefont {Wellstood}}, \ and\ \bibinfo {author}
  {\bibfnamefont {M.}~\bibnamefont {Dreyer}},\ }\bibfield  {title} {\enquote
  {\bibinfo {title} {A 30 {mK}, 13.5 {T} scanning tunneling microscope with two
  independent tips},}\ }\href {\doibase 10.1063/1.4871056} {\bibfield
  {journal} {\bibinfo  {journal} {Rev. Sci. Instrum.}\ }\textbf {\bibinfo
  {volume} {85}},\ \bibinfo {pages} {043706} (\bibinfo {year}
  {2014})}\BibitemShut {NoStop}%
\bibitem [{\citenamefont {Machida}, \citenamefont {Kohsaka},\ and\
  \citenamefont {Hanaguri}(2018)}]{Machida2018}%
  \BibitemOpen
  \bibfield  {author} {\bibinfo {author} {\bibfnamefont {T.}~\bibnamefont
  {Machida}}, \bibinfo {author} {\bibfnamefont {Y.}~\bibnamefont {Kohsaka}}, \
  and\ \bibinfo {author} {\bibfnamefont {T.}~\bibnamefont {Hanaguri}},\
  }\bibfield  {title} {\enquote {\bibinfo {title} {A scanning tunneling
  microscope for spectroscopic imaging below 90 {mK} in magnetic fields up to
  17.5 {T}},}\ }\href {\doibase 10.1063/1.5049619} {\bibfield  {journal}
  {\bibinfo  {journal} {Rev. Sci. Instrum.}\ }\textbf {\bibinfo {volume}
  {89}},\ \bibinfo {pages} {093707} (\bibinfo {year} {2018})}\BibitemShut
  {NoStop}%
\bibitem [{\citenamefont {Balashov}, \citenamefont {Meyer},\ and\ \citenamefont
  {Wulfhekel}(2018)}]{Balashov2018}%
  \BibitemOpen
  \bibfield  {author} {\bibinfo {author} {\bibfnamefont {T.}~\bibnamefont
  {Balashov}}, \bibinfo {author} {\bibfnamefont {M.}~\bibnamefont {Meyer}}, \
  and\ \bibinfo {author} {\bibfnamefont {W.}~\bibnamefont {Wulfhekel}},\
  }\bibfield  {title} {\enquote {\bibinfo {title} {A compact ultrahigh vacuum
  scanning tunneling microscope with dilution refrigeration},}\ }\href
  {\doibase 10.1063/1.5043636} {\bibfield  {journal} {\bibinfo  {journal} {Rev.
  Sci. Instrum.}\ }\textbf {\bibinfo {volume} {89}},\ \bibinfo {pages} {113707}
  (\bibinfo {year} {2018})}\BibitemShut {NoStop}%
\bibitem [{\citenamefont {von Allw\"orden}\ \emph {et~al.}(2018)\citenamefont
  {von Allw\"orden}, \citenamefont {Eich}, \citenamefont {Knol}, \citenamefont
  {Hermenau}, \citenamefont {Sonntag}, \citenamefont {Gerritsen}, \citenamefont
  {Wegner},\ and\ \citenamefont {Khajetoorians}}]{VonAllworden2018}%
  \BibitemOpen
  \bibfield  {author} {\bibinfo {author} {\bibfnamefont {H.}~\bibnamefont {von
  Allw\"orden}}, \bibinfo {author} {\bibfnamefont {A.}~\bibnamefont {Eich}},
  \bibinfo {author} {\bibfnamefont {E.~J.}\ \bibnamefont {Knol}}, \bibinfo
  {author} {\bibfnamefont {J.}~\bibnamefont {Hermenau}}, \bibinfo {author}
  {\bibfnamefont {A.}~\bibnamefont {Sonntag}}, \bibinfo {author} {\bibfnamefont
  {J.~W.}\ \bibnamefont {Gerritsen}}, \bibinfo {author} {\bibfnamefont
  {D.}~\bibnamefont {Wegner}}, \ and\ \bibinfo {author} {\bibfnamefont {A.~A.}\
  \bibnamefont {Khajetoorians}},\ }\bibfield  {title} {\enquote {\bibinfo
  {title} {Design and performance of an ultra-high vacuum spin-polarized
  scanning tunneling microscope operating at 30 {mK} and in a vector magnetic
  field},}\ }\href {\doibase 10.1063/1.5020045} {\bibfield  {journal} {\bibinfo
   {journal} {Rev. Sci. Instrum.}\ }\textbf {\bibinfo {volume} {89}},\ \bibinfo
  {pages} {033902} (\bibinfo {year} {2018})}\BibitemShut {NoStop}%
\bibitem [{\citenamefont {Wong}\ \emph {et~al.}(2020)\citenamefont {Wong},
  \citenamefont {Jeon}, \citenamefont {Nuckolls}, \citenamefont {Oh},
  \citenamefont {Kingsley},\ and\ \citenamefont {Yazdani}}]{Wong2020}%
  \BibitemOpen
  \bibfield  {author} {\bibinfo {author} {\bibfnamefont {D.}~\bibnamefont
  {Wong}}, \bibinfo {author} {\bibfnamefont {S.}~\bibnamefont {Jeon}}, \bibinfo
  {author} {\bibfnamefont {K.~P.}\ \bibnamefont {Nuckolls}}, \bibinfo {author}
  {\bibfnamefont {M.}~\bibnamefont {Oh}}, \bibinfo {author} {\bibfnamefont
  {S.~C.~J.}\ \bibnamefont {Kingsley}}, \ and\ \bibinfo {author} {\bibfnamefont
  {A.}~\bibnamefont {Yazdani}},\ }\bibfield  {title} {\enquote {\bibinfo
  {title} {A modular ultra-high vacuum millikelvin scanning tunneling
  microscope},}\ }\href {\doibase 10.1063/1.5132872} {\bibfield  {journal}
  {\bibinfo  {journal} {Rev. Sci. Instrum.}\ }\textbf {\bibinfo {volume}
  {91}},\ \bibinfo {pages} {023703} (\bibinfo {year} {2020})}\BibitemShut
  {NoStop}%
\bibitem [{\citenamefont {Schwenk}\ \emph {et~al.}(2020)\citenamefont
  {Schwenk}, \citenamefont {Kim}, \citenamefont {Berwanger}, \citenamefont
  {Ghahari}, \citenamefont {Slot}, \citenamefont {Le}, \citenamefont {Cullen},
  \citenamefont {Blankenship}, \citenamefont {Hug}, \citenamefont {Kuk},
  \citenamefont {Giessibl},\ and\ \citenamefont {Stroscio}}]{Schwenk2020}%
  \BibitemOpen
  \bibfield  {author} {\bibinfo {author} {\bibfnamefont {J.}~\bibnamefont
  {Schwenk}}, \bibinfo {author} {\bibfnamefont {S.}~\bibnamefont {Kim}},
  \bibinfo {author} {\bibfnamefont {J.}~\bibnamefont {Berwanger}}, \bibinfo
  {author} {\bibfnamefont {F.}~\bibnamefont {Ghahari}}, \bibinfo {author}
  {\bibfnamefont {M.~R.}\ \bibnamefont {Slot}}, \bibinfo {author}
  {\bibfnamefont {S.~T.}\ \bibnamefont {Le}}, \bibinfo {author} {\bibfnamefont
  {W.~G.}\ \bibnamefont {Cullen}}, \bibinfo {author} {\bibfnamefont {S.~R.}\
  \bibnamefont {Blankenship}}, \bibinfo {author} {\bibfnamefont {H.~J.}\
  \bibnamefont {Hug}}, \bibinfo {author} {\bibfnamefont {Y.}~\bibnamefont
  {Kuk}}, \bibinfo {author} {\bibfnamefont {F.~J.}\ \bibnamefont {Giessibl}}, \
  and\ \bibinfo {author} {\bibfnamefont {J.~A.}\ \bibnamefont {Stroscio}},\
  }\bibfield  {title} {\enquote {\bibinfo {title} {Achieving $\mu${eV}
  tunneling resolution in an in-operando scanning tunneling microscopy, atomic
  force microscopy, and magnetotransport system for quantum materials
  research},}\ }\href {\doibase 10.1063/5.0005320} {\bibfield  {journal}
  {\bibinfo  {journal} {Rev. Sci. Instrum.}\ }\textbf {\bibinfo {volume}
  {91}},\ \bibinfo {pages} {071101} (\bibinfo {year} {2020})}\BibitemShut
  {NoStop}%
\bibitem [{\citenamefont {Esat}\ \emph {et~al.}(2021)\citenamefont {Esat},
  \citenamefont {Borgens}, \citenamefont {Yang}, \citenamefont {Coenen},
  \citenamefont {Cherepanov}, \citenamefont {Raccanelli}, \citenamefont
  {Tautz},\ and\ \citenamefont {Temirov}}]{Esat2021}%
  \BibitemOpen
  \bibfield  {author} {\bibinfo {author} {\bibfnamefont {T.}~\bibnamefont
  {Esat}}, \bibinfo {author} {\bibfnamefont {P.}~\bibnamefont {Borgens}},
  \bibinfo {author} {\bibfnamefont {X.}~\bibnamefont {Yang}}, \bibinfo {author}
  {\bibfnamefont {P.}~\bibnamefont {Coenen}}, \bibinfo {author} {\bibfnamefont
  {V.}~\bibnamefont {Cherepanov}}, \bibinfo {author} {\bibfnamefont
  {A.}~\bibnamefont {Raccanelli}}, \bibinfo {author} {\bibfnamefont {F.~S.}\
  \bibnamefont {Tautz}}, \ and\ \bibinfo {author} {\bibfnamefont
  {R.}~\bibnamefont {Temirov}},\ }\bibfield  {title} {\enquote {\bibinfo
  {title} {A millikelvin scanning tunneling microscope in ultra-high vacuum
  with adiabatic demagnetization refrigeration},}\ }\href {\doibase
  10.1063/5.0050532} {\bibfield  {journal} {\bibinfo  {journal} {Rev. Sci.
  Instrum.}\ }\textbf {\bibinfo {volume} {92}},\ \bibinfo {pages} {063701}
  (\bibinfo {year} {2021})}\BibitemShut {NoStop}%
\bibitem [{\citenamefont {Fern{\'{a}}ndez-Lomana}\ \emph
  {et~al.}(2021)\citenamefont {Fern{\'{a}}ndez-Lomana}, \citenamefont {Wu},
  \citenamefont {Mart{\'{i}}n-Vega}, \citenamefont {S{\'{a}}nchez-Barquilla},
  \citenamefont {{\'{A}}lvarez-Montoya}, \citenamefont {Castilla},
  \citenamefont {Navarrete}, \citenamefont {Marijuan}, \citenamefont {Herrera},
  \citenamefont {Suderow},\ and\ \citenamefont
  {Guillam{\'{o}}n}}]{Fernandez-Lomana2021}%
  \BibitemOpen
  \bibfield  {author} {\bibinfo {author} {\bibfnamefont {M.}~\bibnamefont
  {Fern{\'{a}}ndez-Lomana}}, \bibinfo {author} {\bibfnamefont {B.}~\bibnamefont
  {Wu}}, \bibinfo {author} {\bibfnamefont {F.}~\bibnamefont
  {Mart{\'{i}}n-Vega}}, \bibinfo {author} {\bibfnamefont {R.}~\bibnamefont
  {S{\'{a}}nchez-Barquilla}}, \bibinfo {author} {\bibfnamefont
  {R.}~\bibnamefont {{\'{A}}lvarez-Montoya}}, \bibinfo {author} {\bibfnamefont
  {J.~M.}\ \bibnamefont {Castilla}}, \bibinfo {author} {\bibfnamefont
  {J.}~\bibnamefont {Navarrete}}, \bibinfo {author} {\bibfnamefont {J.~R.}\
  \bibnamefont {Marijuan}}, \bibinfo {author} {\bibfnamefont {E.}~\bibnamefont
  {Herrera}}, \bibinfo {author} {\bibfnamefont {H.}~\bibnamefont {Suderow}}, \
  and\ \bibinfo {author} {\bibfnamefont {I.}~\bibnamefont {Guillam{\'{o}}n}},\
  }\bibfield  {title} {\enquote {\bibinfo {title} {{Millikelvin scanning
  tunneling microscope at 20/22 T with a graphite enabled stick-slip approach
  and an energy resolution below 8 $\mu$eV: Application to conductance
  quantization at 20 T in single atom point contacts of Al and Au and to the
  charge density wave of 2H-NbSe$_2$}},}\ }\href {\doibase 10.1063/5.0059394}
  {\bibfield  {journal} {\bibinfo  {journal} {Rev. Sci. Instrum.}\ }\textbf
  {\bibinfo {volume} {92}},\ \bibinfo {pages} {093701} (\bibinfo {year}
  {2021})}\BibitemShut {NoStop}%
\bibitem [{\citenamefont {Machida}\ \emph {et~al.}(2019)\citenamefont
  {Machida}, \citenamefont {Sun}, \citenamefont {Pyon}, \citenamefont {Takeda},
  \citenamefont {Kohsaka}, \citenamefont {Hanaguri}, \citenamefont {Sasagawa},\
  and\ \citenamefont {Tamegai}}]{Machida2019}%
  \BibitemOpen
  \bibfield  {author} {\bibinfo {author} {\bibfnamefont {T.}~\bibnamefont
  {Machida}}, \bibinfo {author} {\bibfnamefont {Y.}~\bibnamefont {Sun}},
  \bibinfo {author} {\bibfnamefont {S.}~\bibnamefont {Pyon}}, \bibinfo {author}
  {\bibfnamefont {S.}~\bibnamefont {Takeda}}, \bibinfo {author} {\bibfnamefont
  {Y.}~\bibnamefont {Kohsaka}}, \bibinfo {author} {\bibfnamefont
  {T.}~\bibnamefont {Hanaguri}}, \bibinfo {author} {\bibfnamefont
  {T.}~\bibnamefont {Sasagawa}}, \ and\ \bibinfo {author} {\bibfnamefont
  {T.}~\bibnamefont {Tamegai}},\ }\bibfield  {title} {\enquote {\bibinfo
  {title} {Zero-energy vortex bound state in the superconducting topological
  surface state of {Fe(Se,Te)}},}\ }\href {\doibase 10.1038/s41563-019-0397-1}
  {\bibfield  {journal} {\bibinfo  {journal} {Nat. Mater.}\ }\textbf {\bibinfo
  {volume} {18}},\ \bibinfo {pages} {811--815} (\bibinfo {year}
  {2019})}\BibitemShut {NoStop}%
\bibitem [{\citenamefont {Nuckolls}\ \emph {et~al.}(2020)\citenamefont
  {Nuckolls}, \citenamefont {Oh}, \citenamefont {Wong}, \citenamefont {Lian},
  \citenamefont {Watanabe}, \citenamefont {Taniguchi}, \citenamefont
  {Bernevig},\ and\ \citenamefont {Yazdani}}]{Nuckolls2020}%
  \BibitemOpen
  \bibfield  {author} {\bibinfo {author} {\bibfnamefont {K.~P.}\ \bibnamefont
  {Nuckolls}}, \bibinfo {author} {\bibfnamefont {M.}~\bibnamefont {Oh}},
  \bibinfo {author} {\bibfnamefont {D.}~\bibnamefont {Wong}}, \bibinfo {author}
  {\bibfnamefont {B.}~\bibnamefont {Lian}}, \bibinfo {author} {\bibfnamefont
  {K.}~\bibnamefont {Watanabe}}, \bibinfo {author} {\bibfnamefont
  {T.}~\bibnamefont {Taniguchi}}, \bibinfo {author} {\bibfnamefont {B.~A.}\
  \bibnamefont {Bernevig}}, \ and\ \bibinfo {author} {\bibfnamefont
  {A.}~\bibnamefont {Yazdani}},\ }\bibfield  {title} {\enquote {\bibinfo
  {title} {Strongly correlated {Chern} insulators in magic-angle twisted
  bilayer graphene},}\ }\href {\doibase 10.1038/s41586-020-3028-8} {\bibfield
  {journal} {\bibinfo  {journal} {Nature}\ }\textbf {\bibinfo {volume} {588}},\
  \bibinfo {pages} {610--615} (\bibinfo {year} {2020})}\BibitemShut {NoStop}%
\bibitem [{\citenamefont {Kamber}\ \emph {et~al.}(2020)\citenamefont {Kamber},
  \citenamefont {Bergman}, \citenamefont {Eich}, \citenamefont {Iu\c{s}an},
  \citenamefont {Steinbrecher}, \citenamefont {Hauptmann}, \citenamefont
  {Nordstr\"om}, \citenamefont {Katsnelson}, \citenamefont {Wegner},
  \citenamefont {Eriksson},\ and\ \citenamefont {Khajetoorians}}]{Kamber2020}%
  \BibitemOpen
  \bibfield  {author} {\bibinfo {author} {\bibfnamefont {U.}~\bibnamefont
  {Kamber}}, \bibinfo {author} {\bibfnamefont {A.}~\bibnamefont {Bergman}},
  \bibinfo {author} {\bibfnamefont {A.}~\bibnamefont {Eich}}, \bibinfo {author}
  {\bibfnamefont {D.}~\bibnamefont {Iu\c{s}an}}, \bibinfo {author}
  {\bibfnamefont {M.}~\bibnamefont {Steinbrecher}}, \bibinfo {author}
  {\bibfnamefont {N.}~\bibnamefont {Hauptmann}}, \bibinfo {author}
  {\bibfnamefont {L.}~\bibnamefont {Nordstr\"om}}, \bibinfo {author}
  {\bibfnamefont {M.~I.}\ \bibnamefont {Katsnelson}}, \bibinfo {author}
  {\bibfnamefont {D.}~\bibnamefont {Wegner}}, \bibinfo {author} {\bibfnamefont
  {O.}~\bibnamefont {Eriksson}}, \ and\ \bibinfo {author} {\bibfnamefont
  {A.~A.}\ \bibnamefont {Khajetoorians}},\ }\bibfield  {title} {\enquote
  {\bibinfo {title} {Self-induced spin glass state in elemental and crystalline
  neodymium},}\ }\href {\doibase 10.1126/science.aay6757} {\bibfield  {journal}
  {\bibinfo  {journal} {Science}\ }\textbf {\bibinfo {volume} {368}},\ \bibinfo
  {pages} {eaay6757} (\bibinfo {year} {2020})}\BibitemShut {NoStop}%
\bibitem [{\citenamefont {Song}\ \emph
  {et~al.}(2010{\natexlab{b}})\citenamefont {Song}, \citenamefont {Otte},
  \citenamefont {Kuk}, \citenamefont {Hu}, \citenamefont {Torrance},
  \citenamefont {First}, \citenamefont {Heer}, \citenamefont {Min},
  \citenamefont {Adam}, \citenamefont {Stiles}, \citenamefont {MacDonald},\
  and\ \citenamefont {Stroscio}}]{Song2010_1}%
  \BibitemOpen
  \bibfield  {author} {\bibinfo {author} {\bibfnamefont {Y.~J.}\ \bibnamefont
  {Song}}, \bibinfo {author} {\bibfnamefont {A.~F.}\ \bibnamefont {Otte}},
  \bibinfo {author} {\bibfnamefont {Y.}~\bibnamefont {Kuk}}, \bibinfo {author}
  {\bibfnamefont {Y.}~\bibnamefont {Hu}}, \bibinfo {author} {\bibfnamefont
  {D.~B.}\ \bibnamefont {Torrance}}, \bibinfo {author} {\bibfnamefont {P.~N.}\
  \bibnamefont {First}}, \bibinfo {author} {\bibfnamefont {W.~A.~D.}\
  \bibnamefont {Heer}}, \bibinfo {author} {\bibfnamefont {H.}~\bibnamefont
  {Min}}, \bibinfo {author} {\bibfnamefont {S.}~\bibnamefont {Adam}}, \bibinfo
  {author} {\bibfnamefont {M.~D.}\ \bibnamefont {Stiles}}, \bibinfo {author}
  {\bibfnamefont {A.~H.}\ \bibnamefont {MacDonald}}, \ and\ \bibinfo {author}
  {\bibfnamefont {J.~A.}\ \bibnamefont {Stroscio}},\ }\bibfield  {title}
  {\enquote {\bibinfo {title} {High-resolution tunnelling spectroscopy of a
  graphene quartet},}\ }\href {\doibase 10.1038/nature09330} {\bibfield
  {journal} {\bibinfo  {journal} {Nature}\ }\textbf {\bibinfo {volume} {467}},\
  \bibinfo {pages} {185--189} (\bibinfo {year}
  {2010}{\natexlab{b}})}\BibitemShut {NoStop}%
\bibitem [{\citenamefont {Yazdani}, \citenamefont {da~Silva~Neto},\ and\
  \citenamefont {Aynajian}(2016)}]{Yazdani2016}%
  \BibitemOpen
  \bibfield  {author} {\bibinfo {author} {\bibfnamefont {A.}~\bibnamefont
  {Yazdani}}, \bibinfo {author} {\bibfnamefont {E.~H.}\ \bibnamefont
  {da~Silva~Neto}}, \ and\ \bibinfo {author} {\bibfnamefont {P.}~\bibnamefont
  {Aynajian}},\ }\bibfield  {title} {\enquote {\bibinfo {title} {Spectroscopic
  imaging of strongly correlated electronic states},}\ }\href {\doibase
  10.1146/annurev-conmatphys-031214-014529} {\bibfield  {journal} {\bibinfo
  {journal} {Annu. Rev. Conden. Ma. P.}\ }\textbf {\bibinfo {volume} {7}},\
  \bibinfo {pages} {11--33} (\bibinfo {year} {2016})}\BibitemShut {NoStop}%
\bibitem [{\citenamefont {Feldman}\ \emph {et~al.}(2017)\citenamefont
  {Feldman}, \citenamefont {Randeria}, \citenamefont {Li}, \citenamefont
  {Jeon}, \citenamefont {Xie}, \citenamefont {Wang}, \citenamefont {Drozdov},
  \citenamefont {Bernevig},\ and\ \citenamefont {Yazdani}}]{Feldman2017}%
  \BibitemOpen
  \bibfield  {author} {\bibinfo {author} {\bibfnamefont {B.~E.}\ \bibnamefont
  {Feldman}}, \bibinfo {author} {\bibfnamefont {M.~T.}\ \bibnamefont
  {Randeria}}, \bibinfo {author} {\bibfnamefont {J.}~\bibnamefont {Li}},
  \bibinfo {author} {\bibfnamefont {S.}~\bibnamefont {Jeon}}, \bibinfo {author}
  {\bibfnamefont {Y.}~\bibnamefont {Xie}}, \bibinfo {author} {\bibfnamefont
  {Z.}~\bibnamefont {Wang}}, \bibinfo {author} {\bibfnamefont {I.~K.}\
  \bibnamefont {Drozdov}}, \bibinfo {author} {\bibfnamefont {B.~A.}\
  \bibnamefont {Bernevig}}, \ and\ \bibinfo {author} {\bibfnamefont
  {A.}~\bibnamefont {Yazdani}},\ }\bibfield  {title} {\enquote {\bibinfo
  {title} {High-resolution studies of the {Majorana} atomic chain platform},}\
  }\href {\doibase 10.1038/nphys3947} {\bibfield  {journal} {\bibinfo
  {journal} {Nat. Phys.}\ }\textbf {\bibinfo {volume} {13}},\ \bibinfo {pages}
  {286--291} (\bibinfo {year} {2017})}\BibitemShut {NoStop}%
\bibitem [{\citenamefont {Grabert}\ and\ \citenamefont
  {Devoret}(1992)}]{Grabert1992}%
  \BibitemOpen
  \bibinfo {editor} {\bibfnamefont {H.}~\bibnamefont {Grabert}}\ and\ \bibinfo
  {editor} {\bibfnamefont {M.~H.}\ \bibnamefont {Devoret}},\ eds.,\ \href@noop
  {} {\emph {\bibinfo {title} {{Single Charge Tunneling: Coulomb Blockade
  Phenomena in Nanostructures}}}}\ (\bibinfo  {publisher} {Plenum Press},\
  \bibinfo {year} {1992})\BibitemShut {NoStop}%
\bibitem [{\citenamefont {Bardeen}(1961)}]{Bardeen1961}%
  \BibitemOpen
  \bibfield  {author} {\bibinfo {author} {\bibfnamefont {J.}~\bibnamefont
  {Bardeen}},\ }\bibfield  {title} {\enquote {\bibinfo {title} {Tunnelling from
  a many-particle point of view},}\ }\href {\doibase 10.1103/PhysRevLett.6.57}
  {\bibfield  {journal} {\bibinfo  {journal} {Phys. Rev. Lett.}\ }\textbf
  {\bibinfo {volume} {6}},\ \bibinfo {pages} {57--59} (\bibinfo {year}
  {1961})}\BibitemShut {NoStop}%
\bibitem [{\citenamefont {Voigtl\"{a}nder}(2015)}]{Voigtlaender2015}%
  \BibitemOpen
  \bibfield  {author} {\bibinfo {author} {\bibfnamefont {B.}~\bibnamefont
  {Voigtl\"{a}nder}},\ }\href {\doibase 10.1007/978-3-662-45240-0} {\emph
  {\bibinfo {title} {Scanning Probe Microscopy}}}\ (\bibinfo  {publisher}
  {Springer Berlin Heidelberg},\ \bibinfo {year} {2015})\BibitemShut {NoStop}%
\bibitem [{\citenamefont {Gottlieb}\ and\ \citenamefont
  {Wesoloski}(2006)}]{Gottlieb2006}%
  \BibitemOpen
  \bibfield  {author} {\bibinfo {author} {\bibfnamefont {A.~D.}\ \bibnamefont
  {Gottlieb}}\ and\ \bibinfo {author} {\bibfnamefont {L.}~\bibnamefont
  {Wesoloski}},\ }\bibfield  {title} {\enquote {\bibinfo {title} {Bardeen's
  tunnelling theory as applied to scanning tunnelling microscopy: a technical
  guide to the traditional interpretation},}\ }\href {\doibase
  10.1088/0957-4484/17/8/r01} {\bibfield  {journal} {\bibinfo  {journal}
  {Nanotechnology}\ }\textbf {\bibinfo {volume} {17}},\ \bibinfo {pages}
  {R57--R65} (\bibinfo {year} {2006})}\BibitemShut {NoStop}%
\bibitem [{\citenamefont {Tinkham}(1996)}]{Tinkham1996}%
  \BibitemOpen
  \bibfield  {author} {\bibinfo {author} {\bibfnamefont {M.}~\bibnamefont
  {Tinkham}},\ }\href@noop {} {\emph {\bibinfo {title} {{Introduction to
  Supeconductivity}}}}\ (\bibinfo  {publisher} {McGraw Hill},\ \bibinfo {year}
  {1996})\BibitemShut {NoStop}%
\bibitem [{\citenamefont {J{\"{a}}ck}\ \emph {et~al.}(2015)\citenamefont
  {J{\"{a}}ck}, \citenamefont {Eltschka}, \citenamefont {Assig}, \citenamefont
  {Hardock}, \citenamefont {Etzkorn}, \citenamefont {Ast},\ and\ \citenamefont
  {Kern}}]{Jack2015}%
  \BibitemOpen
  \bibfield  {author} {\bibinfo {author} {\bibfnamefont {B.}~\bibnamefont
  {J{\"{a}}ck}}, \bibinfo {author} {\bibfnamefont {M.}~\bibnamefont
  {Eltschka}}, \bibinfo {author} {\bibfnamefont {M.}~\bibnamefont {Assig}},
  \bibinfo {author} {\bibfnamefont {A.}~\bibnamefont {Hardock}}, \bibinfo
  {author} {\bibfnamefont {M.}~\bibnamefont {Etzkorn}}, \bibinfo {author}
  {\bibfnamefont {C.~R.}\ \bibnamefont {Ast}}, \ and\ \bibinfo {author}
  {\bibfnamefont {K.}~\bibnamefont {Kern}},\ }\bibfield  {title} {\enquote
  {\bibinfo {title} {{A nanoscale gigahertz source realized with Josephson
  scanning tunneling microscopy}},}\ }\href {\doibase 10.1063/1.4905322}
  {\bibfield  {journal} {\bibinfo  {journal} {Appl. Phys. Lett.}\ }\textbf
  {\bibinfo {volume} {106}},\ \bibinfo {pages} {013109} (\bibinfo {year}
  {2015})}\BibitemShut {NoStop}%
\bibitem [{\citenamefont {Ast}\ \emph {et~al.}(2016)\citenamefont {Ast},
  \citenamefont {J{\"{a}}ck}, \citenamefont {Senkpiel}, \citenamefont
  {Eltschka}, \citenamefont {Etzkorn}, \citenamefont {Ankerhold},\ and\
  \citenamefont {Kern}}]{Ast2016}%
  \BibitemOpen
  \bibfield  {author} {\bibinfo {author} {\bibfnamefont {C.~R.}\ \bibnamefont
  {Ast}}, \bibinfo {author} {\bibfnamefont {B.}~\bibnamefont {J{\"{a}}ck}},
  \bibinfo {author} {\bibfnamefont {J.}~\bibnamefont {Senkpiel}}, \bibinfo
  {author} {\bibfnamefont {M.}~\bibnamefont {Eltschka}}, \bibinfo {author}
  {\bibfnamefont {M.}~\bibnamefont {Etzkorn}}, \bibinfo {author} {\bibfnamefont
  {J.}~\bibnamefont {Ankerhold}}, \ and\ \bibinfo {author} {\bibfnamefont
  {K.}~\bibnamefont {Kern}},\ }\bibfield  {title} {\enquote {\bibinfo {title}
  {{Sensing the quantum limit in scanning tunnelling spectroscopy}},}\ }\href
  {\doibase 10.1038/ncomms13009} {\bibfield  {journal} {\bibinfo  {journal}
  {Nat. Commun.}\ }\textbf {\bibinfo {volume} {7}},\ \bibinfo {pages} {13009}
  (\bibinfo {year} {2016})}\BibitemShut {NoStop}%
\bibitem [{\citenamefont {J\"ack}\ \emph {et~al.}(2016)\citenamefont {J\"ack},
  \citenamefont {Eltschka}, \citenamefont {Assig}, \citenamefont {Etzkorn},
  \citenamefont {Ast},\ and\ \citenamefont {Kern}}]{Jack2016}%
  \BibitemOpen
  \bibfield  {author} {\bibinfo {author} {\bibfnamefont {B.}~\bibnamefont
  {J\"ack}}, \bibinfo {author} {\bibfnamefont {M.}~\bibnamefont {Eltschka}},
  \bibinfo {author} {\bibfnamefont {M.}~\bibnamefont {Assig}}, \bibinfo
  {author} {\bibfnamefont {M.}~\bibnamefont {Etzkorn}}, \bibinfo {author}
  {\bibfnamefont {C.~R.}\ \bibnamefont {Ast}}, \ and\ \bibinfo {author}
  {\bibfnamefont {K.}~\bibnamefont {Kern}},\ }\bibfield  {title} {\enquote
  {\bibinfo {title} {Critical {Josephson} current in the dynamical {Coulomb}
  blockade regime},}\ }\href {\doibase 10.1103/PhysRevB.93.020504} {\bibfield
  {journal} {\bibinfo  {journal} {Phys. Rev. B}\ }\textbf {\bibinfo {volume}
  {93}},\ \bibinfo {pages} {020504} (\bibinfo {year} {2016})}\BibitemShut
  {NoStop}%
\bibitem [{\citenamefont {J\"ack}\ \emph {et~al.}(2017)\citenamefont {J\"ack},
  \citenamefont {Senkpiel}, \citenamefont {Etzkorn}, \citenamefont {Ankerhold},
  \citenamefont {Ast},\ and\ \citenamefont {Kern}}]{Jack2017}%
  \BibitemOpen
  \bibfield  {author} {\bibinfo {author} {\bibfnamefont {B.}~\bibnamefont
  {J\"ack}}, \bibinfo {author} {\bibfnamefont {J.}~\bibnamefont {Senkpiel}},
  \bibinfo {author} {\bibfnamefont {M.}~\bibnamefont {Etzkorn}}, \bibinfo
  {author} {\bibfnamefont {J.}~\bibnamefont {Ankerhold}}, \bibinfo {author}
  {\bibfnamefont {C.~R.}\ \bibnamefont {Ast}}, \ and\ \bibinfo {author}
  {\bibfnamefont {K.}~\bibnamefont {Kern}},\ }\bibfield  {title} {\enquote
  {\bibinfo {title} {Quantum {Brownian} motion at strong dissipation probed by
  superconducting tunnel junctions},}\ }\href {\doibase
  10.1103/PhysRevLett.119.147702} {\bibfield  {journal} {\bibinfo  {journal}
  {Phys. Rev. Lett.}\ }\textbf {\bibinfo {volume} {119}},\ \bibinfo {pages}
  {147702} (\bibinfo {year} {2017})}\BibitemShut {NoStop}%
\bibitem [{\citenamefont {Senkpiel}\ \emph
  {et~al.}(2020{\natexlab{a}})\citenamefont {Senkpiel}, \citenamefont
  {Dambach}, \citenamefont {Etzkorn}, \citenamefont {Drost}, \citenamefont
  {Padurariu}, \citenamefont {Kubala}, \citenamefont {Belzig}, \citenamefont
  {Yeyati}, \citenamefont {Cuevas}, \citenamefont {Ankerhold}, \citenamefont
  {Ast},\ and\ \citenamefont {Kern}}]{Senkpiel2020_1}%
  \BibitemOpen
  \bibfield  {author} {\bibinfo {author} {\bibfnamefont {J.}~\bibnamefont
  {Senkpiel}}, \bibinfo {author} {\bibfnamefont {S.}~\bibnamefont {Dambach}},
  \bibinfo {author} {\bibfnamefont {M.}~\bibnamefont {Etzkorn}}, \bibinfo
  {author} {\bibfnamefont {R.}~\bibnamefont {Drost}}, \bibinfo {author}
  {\bibfnamefont {C.}~\bibnamefont {Padurariu}}, \bibinfo {author}
  {\bibfnamefont {B.}~\bibnamefont {Kubala}}, \bibinfo {author} {\bibfnamefont
  {W.}~\bibnamefont {Belzig}}, \bibinfo {author} {\bibfnamefont {A.~L.}\
  \bibnamefont {Yeyati}}, \bibinfo {author} {\bibfnamefont {J.~C.}\
  \bibnamefont {Cuevas}}, \bibinfo {author} {\bibfnamefont {J.}~\bibnamefont
  {Ankerhold}}, \bibinfo {author} {\bibfnamefont {C.~R.}\ \bibnamefont {Ast}},
  \ and\ \bibinfo {author} {\bibfnamefont {K.}~\bibnamefont {Kern}},\
  }\bibfield  {title} {\enquote {\bibinfo {title} {Single channel {Josephson}
  effect in a high transmission atomic contact},}\ }\href {\doibase
  10.1038/s42005-020-00397-z} {\bibfield  {journal} {\bibinfo  {journal}
  {Commun. Phys.}\ }\textbf {\bibinfo {volume} {3}},\ \bibinfo {pages} {131}
  (\bibinfo {year} {2020}{\natexlab{a}})}\BibitemShut {NoStop}%
\bibitem [{\citenamefont {Senkpiel}\ \emph
  {et~al.}(2020{\natexlab{b}})\citenamefont {Senkpiel}, \citenamefont
  {Kl{\"{o}}ckner}, \citenamefont {Etzkorn}, \citenamefont {Dambach},
  \citenamefont {Kubala}, \citenamefont {Belzig}, \citenamefont {Yeyati},
  \citenamefont {Cuevas}, \citenamefont {Pauly}, \citenamefont {Ankerhold},
  \citenamefont {Ast},\ and\ \citenamefont {Kern}}]{Senkpiel2020}%
  \BibitemOpen
  \bibfield  {author} {\bibinfo {author} {\bibfnamefont {J.}~\bibnamefont
  {Senkpiel}}, \bibinfo {author} {\bibfnamefont {J.~C.}\ \bibnamefont
  {Kl{\"{o}}ckner}}, \bibinfo {author} {\bibfnamefont {M.}~\bibnamefont
  {Etzkorn}}, \bibinfo {author} {\bibfnamefont {S.}~\bibnamefont {Dambach}},
  \bibinfo {author} {\bibfnamefont {B.}~\bibnamefont {Kubala}}, \bibinfo
  {author} {\bibfnamefont {W.}~\bibnamefont {Belzig}}, \bibinfo {author}
  {\bibfnamefont {A.~L.}\ \bibnamefont {Yeyati}}, \bibinfo {author}
  {\bibfnamefont {J.~C.}\ \bibnamefont {Cuevas}}, \bibinfo {author}
  {\bibfnamefont {F.}~\bibnamefont {Pauly}}, \bibinfo {author} {\bibfnamefont
  {J.}~\bibnamefont {Ankerhold}}, \bibinfo {author} {\bibfnamefont {C.~R.}\
  \bibnamefont {Ast}}, \ and\ \bibinfo {author} {\bibfnamefont
  {K.}~\bibnamefont {Kern}},\ }\bibfield  {title} {\enquote {\bibinfo {title}
  {Dynamical {Coulomb} blockade as a local probe for quantum transport},}\
  }\href {\doibase 10.1103/PhysRevLett.124.156803} {\bibfield  {journal}
  {\bibinfo  {journal} {Phys. Rev. Lett.}\ }\textbf {\bibinfo {volume} {124}},\
  \bibinfo {pages} {156803} (\bibinfo {year} {2020}{\natexlab{b}})}\BibitemShut
  {NoStop}%
\bibitem [{\citenamefont {Senkpiel}\ \emph {et~al.}(2022)\citenamefont
  {Senkpiel}, \citenamefont {Drost}, \citenamefont {Kl\"ockner}, \citenamefont
  {Etzkorn}, \citenamefont {Ankerhold}, \citenamefont {Cuevas}, \citenamefont
  {Pauly}, \citenamefont {Kern},\ and\ \citenamefont {Ast}}]{Senkpiel2022}%
  \BibitemOpen
  \bibfield  {author} {\bibinfo {author} {\bibfnamefont {J.}~\bibnamefont
  {Senkpiel}}, \bibinfo {author} {\bibfnamefont {R.}~\bibnamefont {Drost}},
  \bibinfo {author} {\bibfnamefont {J.~C.}\ \bibnamefont {Kl\"ockner}},
  \bibinfo {author} {\bibfnamefont {M.}~\bibnamefont {Etzkorn}}, \bibinfo
  {author} {\bibfnamefont {J.}~\bibnamefont {Ankerhold}}, \bibinfo {author}
  {\bibfnamefont {J.~C.}\ \bibnamefont {Cuevas}}, \bibinfo {author}
  {\bibfnamefont {F.}~\bibnamefont {Pauly}}, \bibinfo {author} {\bibfnamefont
  {K.}~\bibnamefont {Kern}}, \ and\ \bibinfo {author} {\bibfnamefont {C.~R.}\
  \bibnamefont {Ast}},\ }\bibfield  {title} {\enquote {\bibinfo {title}
  {Extracting transport channel transmissions in scanning tunneling microscopy
  using superconducting excess current},}\ }\href {\doibase
  10.1103/PhysRevB.105.165401} {\bibfield  {journal} {\bibinfo  {journal}
  {Phys. Rev. B}\ }\textbf {\bibinfo {volume} {105}},\ \bibinfo {pages}
  {165401} (\bibinfo {year} {2022})}\BibitemShut {NoStop}%
\bibitem [{\citenamefont {Martinis}\ and\ \citenamefont
  {Nahum}(1993)}]{Martinis1993}%
  \BibitemOpen
  \bibfield  {author} {\bibinfo {author} {\bibfnamefont {J.~M.}\ \bibnamefont
  {Martinis}}\ and\ \bibinfo {author} {\bibfnamefont {M.}~\bibnamefont
  {Nahum}},\ }\bibfield  {title} {\enquote {\bibinfo {title} {Effect of
  environmental noise on the accuracy of {Coulomb}-blockade devices},}\ }\href
  {\doibase 10.1103/PhysRevB.48.18316} {\bibfield  {journal} {\bibinfo
  {journal} {Phys. Rev. B}\ }\textbf {\bibinfo {volume} {48}},\ \bibinfo
  {pages} {18316--18319} (\bibinfo {year} {1993})}\BibitemShut {NoStop}%
\bibitem [{\citenamefont {Siewert}, \citenamefont {Nazarov},\ and\
  \citenamefont {Falci}(1997)}]{Siewert1997}%
  \BibitemOpen
  \bibfield  {author} {\bibinfo {author} {\bibfnamefont {J.}~\bibnamefont
  {Siewert}}, \bibinfo {author} {\bibfnamefont {Y.~V.}\ \bibnamefont
  {Nazarov}}, \ and\ \bibinfo {author} {\bibfnamefont {G.}~\bibnamefont
  {Falci}},\ }\bibfield  {title} {\enquote {\bibinfo {title} {A generalized
  model of non-thermal noise in the electromagnetic environment of
  small-capacitance tunnel junctions},}\ }\href {\doibase
  10.1209/epl/i1997-00252-0} {\bibfield  {journal} {\bibinfo  {journal}
  {Europhys. Lett.}\ }\textbf {\bibinfo {volume} {38}},\ \bibinfo {pages}
  {365--370} (\bibinfo {year} {1997})}\BibitemShut {NoStop}%
\bibitem [{BEM()}]{BEM}%
  \BibitemOpen
  \href@noop {} {}\bibinfo {note}
  {{https://www.integratedsoft.com/products/Electro}}\BibitemShut {NoStop}%
\bibitem [{\citenamefont {Hergenrother}\ \emph {et~al.}(1994)\citenamefont
  {Hergenrother}, \citenamefont {Tuominen}, \citenamefont {Lu}, \citenamefont
  {Ralph},\ and\ \citenamefont {Tinkham}}]{Hergenrother1994}%
  \BibitemOpen
  \bibfield  {author} {\bibinfo {author} {\bibfnamefont {J.~M.}\ \bibnamefont
  {Hergenrother}}, \bibinfo {author} {\bibfnamefont {M.~T.}\ \bibnamefont
  {Tuominen}}, \bibinfo {author} {\bibfnamefont {J.~G.}\ \bibnamefont {Lu}},
  \bibinfo {author} {\bibfnamefont {D.~C.}\ \bibnamefont {Ralph}}, \ and\
  \bibinfo {author} {\bibfnamefont {M.}~\bibnamefont {Tinkham}},\ }\bibfield
  {title} {\enquote {\bibinfo {title} {Charge transport and photon-assisted
  tunneling in the {NSN} single-electron transistor},}\ }\href {\doibase
  10.1016/0921-4526(94)90077-9} {\bibfield  {journal} {\bibinfo  {journal}
  {Physica B Condens. Matter}\ }\textbf {\bibinfo {volume} {203}},\ \bibinfo
  {pages} {327--339} (\bibinfo {year} {1994})}\BibitemShut {NoStop}%
\bibitem [{\citenamefont {Hergenrother}\ \emph {et~al.}(1995)\citenamefont
  {Hergenrother}, \citenamefont {Lu}, \citenamefont {Tuominen}, \citenamefont
  {Ralph},\ and\ \citenamefont {Tinkham}}]{Hergenrother1995}%
  \BibitemOpen
  \bibfield  {author} {\bibinfo {author} {\bibfnamefont {J.~M.}\ \bibnamefont
  {Hergenrother}}, \bibinfo {author} {\bibfnamefont {J.~G.}\ \bibnamefont
  {Lu}}, \bibinfo {author} {\bibfnamefont {M.~T.}\ \bibnamefont {Tuominen}},
  \bibinfo {author} {\bibfnamefont {D.~C.}\ \bibnamefont {Ralph}}, \ and\
  \bibinfo {author} {\bibfnamefont {M.}~\bibnamefont {Tinkham}},\ }\bibfield
  {title} {\enquote {\bibinfo {title} {Photon-activated switch behavior in the
  single-electron transistor with a superconducting island},}\ }\href {\doibase
  10.1103/PhysRevB.51.9407} {\bibfield  {journal} {\bibinfo  {journal} {Phys.
  Rev. B}\ }\textbf {\bibinfo {volume} {51}},\ \bibinfo {pages} {9407--9410}
  (\bibinfo {year} {1995})}\BibitemShut {NoStop}%
\bibitem [{\citenamefont {Ambegaokar}\ and\ \citenamefont
  {Baratoff}(1963)}]{Ambegaokar1963}%
  \BibitemOpen
  \bibfield  {author} {\bibinfo {author} {\bibfnamefont {V.}~\bibnamefont
  {Ambegaokar}}\ and\ \bibinfo {author} {\bibfnamefont {A.}~\bibnamefont
  {Baratoff}},\ }\bibfield  {title} {\enquote {\bibinfo {title} {{Tunneling
  between superconductors}},}\ }\href {\doibase 10.1103/PhysRevLett.11.104}
  {\bibfield  {journal} {\bibinfo  {journal} {Phys. Rev. Lett.}\ }\textbf
  {\bibinfo {volume} {11}},\ \bibinfo {pages} {104} (\bibinfo {year}
  {1963})}\BibitemShut {NoStop}%
\bibitem [{\citenamefont {Joyez}\ \emph {et~al.}(1999)\citenamefont {Joyez},
  \citenamefont {Vion}, \citenamefont {Götz}, \citenamefont {Devoret},\ and\
  \citenamefont {Esteve}}]{Joyez1999}%
  \BibitemOpen
  \bibfield  {author} {\bibinfo {author} {\bibfnamefont {P.}~\bibnamefont
  {Joyez}}, \bibinfo {author} {\bibfnamefont {D.}~\bibnamefont {Vion}},
  \bibinfo {author} {\bibfnamefont {M.}~\bibnamefont {Götz}}, \bibinfo {author}
  {\bibfnamefont {M.}~\bibnamefont {Devoret}}, \ and\ \bibinfo {author}
  {\bibfnamefont {D.}~\bibnamefont {Esteve}},\ }\bibfield  {title} {\enquote
  {\bibinfo {title} {The josephson effect in nanoscale tunnel junctions},}\
  }\href@noop {} {\bibfield  {journal} {\bibinfo  {journal} {Journal of
  Superconductivity}\ }\textbf {\bibinfo {volume} {12}},\ \bibinfo {pages}
  {757--766} (\bibinfo {year} {1999})}\BibitemShut {NoStop}%
\bibitem [{\citenamefont {Ingold}\ and\ \citenamefont
  {Grabert}(1991)}]{Ingold1991}%
  \BibitemOpen
  \bibfield  {author} {\bibinfo {author} {\bibfnamefont {G.~L.}\ \bibnamefont
  {Ingold}}\ and\ \bibinfo {author} {\bibfnamefont {H.}~\bibnamefont
  {Grabert}},\ }\bibfield  {title} {\enquote {\bibinfo {title}
  {{Finite-temperature current-voltage characteristics of ultrasmall tunnel
  junctions}},}\ }\href {\doibase 10.1209/0295-5075/14/4/015} {\bibfield
  {journal} {\bibinfo  {journal} {Europhys. Lett.}\ }\textbf {\bibinfo {volume}
  {14}},\ \bibinfo {pages} {371--376} (\bibinfo {year} {1991})}\BibitemShut
  {NoStop}%
\bibitem [{\citenamefont {Dynes}, \citenamefont {Narayanamurti},\ and\
  \citenamefont {Garno}(1978)}]{Dynes1978}%
  \BibitemOpen
  \bibfield  {author} {\bibinfo {author} {\bibfnamefont {R.~C.}\ \bibnamefont
  {Dynes}}, \bibinfo {author} {\bibfnamefont {V.}~\bibnamefont
  {Narayanamurti}}, \ and\ \bibinfo {author} {\bibfnamefont {J.~P.}\
  \bibnamefont {Garno}},\ }\bibfield  {title} {\enquote {\bibinfo {title}
  {{Direct measurement of quasiparticle-lifetime broadening in a strong-coupled
  superconductor}},}\ }\href {\doibase 10.1103/PhysRevLett.41.1509} {\bibfield
  {journal} {\bibinfo  {journal} {Phys. Rev. Lett.}\ }\textbf {\bibinfo
  {volume} {41}},\ \bibinfo {pages} {1509--1512} (\bibinfo {year}
  {1978})}\BibitemShut {NoStop}%
\bibitem [{\citenamefont {Maki}(1964)}]{Maki1964}%
  \BibitemOpen
  \bibfield  {author} {\bibinfo {author} {\bibfnamefont {K.}~\bibnamefont
  {Maki}},\ }\bibfield  {title} {\enquote {\bibinfo {title} {{Pauli
  Paramagnetism and Superconducting State. II}},}\ }\href {\doibase
  10.1143/PTP.32.29} {\bibfield  {journal} {\bibinfo  {journal} {Prog. Theor.
  Phys.}\ }\textbf {\bibinfo {volume} {32}},\ \bibinfo {pages} {29--36}
  (\bibinfo {year} {1964})}\BibitemShut {NoStop}%
\bibitem [{\citenamefont {Pekola}\ \emph {et~al.}(2010)\citenamefont {Pekola},
  \citenamefont {Maisi}, \citenamefont {Kafanov}, \citenamefont {Chekurov},
  \citenamefont {Kemppinen}, \citenamefont {Pashkin}, \citenamefont {Saira},
  \citenamefont {M\"ott\"onen},\ and\ \citenamefont {Tsai}}]{Pekola2010}%
  \BibitemOpen
  \bibfield  {author} {\bibinfo {author} {\bibfnamefont {J.~P.}\ \bibnamefont
  {Pekola}}, \bibinfo {author} {\bibfnamefont {V.~F.}\ \bibnamefont {Maisi}},
  \bibinfo {author} {\bibfnamefont {S.}~\bibnamefont {Kafanov}}, \bibinfo
  {author} {\bibfnamefont {N.}~\bibnamefont {Chekurov}}, \bibinfo {author}
  {\bibfnamefont {A.}~\bibnamefont {Kemppinen}}, \bibinfo {author}
  {\bibfnamefont {Y.~A.}\ \bibnamefont {Pashkin}}, \bibinfo {author}
  {\bibfnamefont {O.-P.}\ \bibnamefont {Saira}}, \bibinfo {author}
  {\bibfnamefont {M.}~\bibnamefont {M\"ott\"onen}}, \ and\ \bibinfo {author}
  {\bibfnamefont {J.~S.}\ \bibnamefont {Tsai}},\ }\bibfield  {title} {\enquote
  {\bibinfo {title} {Environment-assisted tunneling as an origin of the {Dynes}
  density of states},}\ }\href {\doibase 10.1103/PhysRevLett.105.026803}
  {\bibfield  {journal} {\bibinfo  {journal} {Phys. Rev. Lett.}\ }\textbf
  {\bibinfo {volume} {105}},\ \bibinfo {pages} {026803} (\bibinfo {year}
  {2010})}\BibitemShut {NoStop}%
\bibitem [{\citenamefont {Nahum}, \citenamefont {Eiles},\ and\ \citenamefont
  {Martinis}(1994)}]{Nahum1994}%
  \BibitemOpen
  \bibfield  {author} {\bibinfo {author} {\bibfnamefont {M.}~\bibnamefont
  {Nahum}}, \bibinfo {author} {\bibfnamefont {T.~M.}\ \bibnamefont {Eiles}}, \
  and\ \bibinfo {author} {\bibfnamefont {J.~M.}\ \bibnamefont {Martinis}},\
  }\bibfield  {title} {\enquote {\bibinfo {title} {{Electronic
  microrefrigerator based on a normal-insulator-superconductor tunnel
  junction}},}\ }\href {\doibase 10.1063/1.112456} {\bibfield  {journal}
  {\bibinfo  {journal} {Appl. Phys. Lett.}\ }\textbf {\bibinfo {volume} {65}},\
  \bibinfo {pages} {3123--3125} (\bibinfo {year} {1994})}\BibitemShut {NoStop}%
\bibitem [{\citenamefont {Gasparovic}, \citenamefont {Taylor},\ and\
  \citenamefont {Eck}(1966)}]{Gasparovic1966}%
  \BibitemOpen
  \bibfield  {author} {\bibinfo {author} {\bibfnamefont {R.~F.}\ \bibnamefont
  {Gasparovic}}, \bibinfo {author} {\bibfnamefont {B.~N.}\ \bibnamefont
  {Taylor}}, \ and\ \bibinfo {author} {\bibfnamefont {R.~E.}\ \bibnamefont
  {Eck}},\ }\bibfield  {title} {\enquote {\bibinfo {title} {Temperature
  dependence of the superconducting energy gap of {Pb}},}\ }\href {\doibase
  10.1016/0038-1098(66)90106-2} {\bibfield  {journal} {\bibinfo  {journal}
  {Solid State Commun.}\ }\textbf {\bibinfo {volume} {4}},\ \bibinfo {pages}
  {59--63} (\bibinfo {year} {1966})}\BibitemShut {NoStop}%
\bibitem [{\citenamefont {Di~Marco}\ \emph {et~al.}(2013)\citenamefont
  {Di~Marco}, \citenamefont {Maisi}, \citenamefont {Pekola},\ and\
  \citenamefont {Hekking}}]{DiMarco2013}%
  \BibitemOpen
  \bibfield  {author} {\bibinfo {author} {\bibfnamefont {A.}~\bibnamefont
  {Di~Marco}}, \bibinfo {author} {\bibfnamefont {V.~F.}\ \bibnamefont {Maisi}},
  \bibinfo {author} {\bibfnamefont {J.~P.}\ \bibnamefont {Pekola}}, \ and\
  \bibinfo {author} {\bibfnamefont {F.~W.~J.}\ \bibnamefont {Hekking}},\
  }\bibfield  {title} {\enquote {\bibinfo {title} {Leakage current of a
  superconductor--normal metal tunnel junction connected to a high-temperature
  environment},}\ }\href {\doibase 10.1103/PhysRevB.88.174507} {\bibfield
  {journal} {\bibinfo  {journal} {Phys. Rev. B}\ }\textbf {\bibinfo {volume}
  {88}},\ \bibinfo {pages} {174507} (\bibinfo {year} {2013})}\BibitemShut
  {NoStop}%
\bibitem [{NF()}]{NF}%
  \BibitemOpen
  \href@noop {} {}\bibinfo {note} {{Model SA-607F2, NF Corporation, 6-3-20
  Tsunashima Higashi, Kohoku-ku, Yokohama 223-8508, Japan}}\BibitemShut
  {NoStop}%
\bibitem [{\citenamefont {Clerk}\ \emph {et~al.}(2010)\citenamefont {Clerk},
  \citenamefont {Devoret}, \citenamefont {Girvin}, \citenamefont {Marquardt},\
  and\ \citenamefont {Schoelkopf}}]{Clerk2010}%
  \BibitemOpen
  \bibfield  {author} {\bibinfo {author} {\bibfnamefont {A.~A.}\ \bibnamefont
  {Clerk}}, \bibinfo {author} {\bibfnamefont {M.~H.}\ \bibnamefont {Devoret}},
  \bibinfo {author} {\bibfnamefont {S.~M.}\ \bibnamefont {Girvin}}, \bibinfo
  {author} {\bibfnamefont {F.}~\bibnamefont {Marquardt}}, \ and\ \bibinfo
  {author} {\bibfnamefont {R.~J.}\ \bibnamefont {Schoelkopf}},\ }\bibfield
  {title} {\enquote {\bibinfo {title} {Introduction to quantum noise,
  measurement, and amplification},}\ }\href {\doibase
  10.1103/RevModPhys.82.1155} {\bibfield  {journal} {\bibinfo  {journal} {Rev.
  Mod. Phys.}\ }\textbf {\bibinfo {volume} {82}},\ \bibinfo {pages}
  {1155--1208} (\bibinfo {year} {2010})}\BibitemShut {NoStop}%
  
\end{thebibliography}
\end {document}